\documentclass{aa}  

\usepackage{graphicx}
\usepackage{xcolor}
\usepackage{booktabs}
\usepackage{supertabular}
\usepackage[version=4]{mhchem}
\usepackage{ulem}
\usepackage{longtable}
\usepackage{caption}
\usepackage{sidecap}
\usepackage{float}
\usepackage{placeins}
\usepackage{rotating}
\usepackage{txfonts}
\usepackage{hyperref}
\hypersetup{
    colorlinks=true,
    linkcolor= blue,
    filecolor=magenta,      
    urlcolor=blue,
    citecolor= blue,
}%
\begin{document}

   \title{JOYS+: Analyses of OCN$^-$, N$_2$O, NO, and complex cyanides in ices}

   \subtitle{Thermal processing results in modest enhancement of OCN$^-$ ice}

   \author{P. Nazari
          \inst{1}
          \and
          N. Brunken\inst{2}
          \and
          Y. Chen\inst{2}  
          \and
          K. Slavicinska\inst{2}
          \and
          E. F. van Dishoeck\inst{2,3}
          \and
          W. R. M. Rocha\inst{2}
          \and
          A. C. A. Boogert\inst{4}
          \and
          M. G. Navarro\inst{5}
          \and
          V. J. M. Le Gouellec\inst{6,7}
          \and
          L. Francis\inst{2}
          \and
          \L. Tychoniec\inst{2}
          \and
          A. Caratti o Garatti\inst{8}
          \and
          C. Gieser\inst{9}
          \and
          T. P. Greene\inst{10}
          \and
          P. J. Kavanagh\inst{11}          
          }

   \institute{European Southern Observatory, Karl-Schwarzschild-Strasse 2, 85748 Garching, Germany\\ 
        \email{pooneh.nazari@eso.org}
         \and
         Leiden Observatory, Leiden University, P.O. Box 9513, 2300 RA Leiden, Netherlands
         \and
         Max Planck Institut f\"{u}r Extraterrestrische Physik (MPE), Giessenbachstrasse 1, 85748 Garching, Germany
         \and
         Institute for Astronomy, University of Hawai’i at Manoa, 2680 Woodlawn Drive, Honolulu, HI 96822, USA
         \and
         INAF–Osservatorio Astronomico di Roma, Via di Frascati 33, 00078 Monte Porzio Catone, Italy
         \and
         Institut de Cienci\`{e}s de l'Espai (ICE-CSIC), Campus UAB, Carrer de Can Magrans S/N, E-08193 Cerdanyola del Vall\`{e}s, Spain
         \and
         Institut d'Estudis Espacials de Catalunya (IEEC), c/ Gran Capit{\'a}, 2–4, 08034 Barcelona, Spain
         \and
         INAF–Osservatorio Astronomico di Capodimonte, Salita Moiariello 16, 80131 Napoli, Italy
         \and
         Max Planck Institute for Astronomy, K{\"o}nigstuhl 17, 69117 Heidelberg, Germany
         \and
         IPAC, California Institute of Technology, MC 100-22, 1200 E. California Blvd., Pasadena, CA, 91125, USA
         \and 
         Department of Physics, Maynooth University, Maynooth, Co. 13 Kildare, Ireland
             }

   \date{Received 27 October 2025/ Accepted 22 April 2026}
  
  \abstract{Nitrogen-bearing molecules are generally more difficult to observe than oxygen-bearing ones in both gas and ice, mainly due to the lower abundance of nitrogen in the interstellar medium. Therefore, the formation pathways of many of these species is still under debate.}
  {Studies prior to the launch of the \textit{James Webb Space Telescope} (JWST) did not have the sensitivity to observe ices toward the youngest and most deeply embedded Class 0 objects. The time is now ripe to study nitrogen-bearing molecules toward ice-rich Class 0 and Class I objects with the JWST. Here we will focus on OCN$^-$, CH$_3$CN, C$_2$H$_5$CN, NO, and N$_2$O in ices to better understand their formation.}
  {We use the data from JWST Observations of Young protoStars (JOYS+) program. Particularly, we study the objects that have the JWST NIRSpec-IFU observations (8 Class 0 and 11 Class I) to measure the ice column densities of the targeted molecules.}
  {We firmly detect OCN$^-$ in ices for all these objects, tentatively detect CH$_3$CN, C$_2$H$_5$CN, and N$_2$O toward three sources, and find upper limits on the NO abundance in ices. The OCN$^-$/CO$_2$ ratios are found to be larger by a factor of ${\sim}2-3$ for the objects that have a visible CO$_2$ (15.2\,$\mu$m) double peak (a sign of ice thermal processing) pointing to the moderate effect of temperature on OCN$^-$ production. Considering H$_2$O, CO$_2$, and OCN$^-$ relations with $A_{\rm V}$, we tentatively find that OCN$^-$ forms at a later stage compared to H$_2$O and CO$_2$. We find that the ratios of CH$_3$CN, C$_2$H$_5$CN, and N$_2$O with respect to OCN$^-$ are relatively constant within one order of magnitude across our objects, likely suggesting that they have similar ice environments. The upper limit abundances of NO are around one order of magnitude lower than what was previously predicted in ices of a mature protoplanetary disk to explain the gas-phase detection of this molecule in the disk. This indicates that gas-phase NO may be a product of another molecule such as N$_2$O in the ices.}
  {We conclude that after OCN$^-$ forms, it can get enhanced at higher temperatures by only a factor of ${\sim}2-3$ and thus OCN$^-$ detection alone does not imply ice heating. Large-sample studies of OCN$^-$ toward pre-stellar cores will be useful to further confirm the formation timeline of this molecule.}


   \maketitle
%

\section{Introduction}

Compared to oxygen, nitrogen is a factor of ${\sim}5$ less abundant in the interstellar medium (ISM, \citealt{Wilson1994}), therefore, molecules containing nitrogen are also relatively less abundant and generally more difficult to observe in the ice and gas. However, understanding the formation of nitrogen-bearing species is crucial considering their contribution to formation of amino acids and nucleobases which are necessary to the formation of habitable worlds. In ices, OCN$^-$ is one of the nitrogen-bearing molecules along with NH$_3$ and NH$_4^+$ that often shows a clear and strong absorption feature (e.g., \citealt{Schutte2003}; \citealt{Broekhuizen2004}; \citealt{Bottinelli2010}; \citealt{Boogert2008, Boogert2015}). Telescopes, especially those from the ground, such as ESO's Very Large Telescope (VLT) and NASA's InfraRed Telescope Facility (IRTF), have observed OCN$^-$ toward low- and high-mass protostellar systems (\citealt{Broekhuizen2005}; \citealt{Boogert2022}). However, for low-mass objects, those observations were biased toward the brighter, less embedded, and more evolved objects (i.e., Class I sources) rather than the younger and more embedded ice-rich ones (i.e., Class 0 sources). With the launch of the \textit{James Webb Space Telescope} (JWST, \citealt{Gardner2023}), we have a unique opportunity to also search for OCN$^-$ and other molecules in ices of the fainter Class 0 objects (\citealt{Yang2022}) and dense pre-stellar cores (\citealt{McClure2023}). 

One of the clearest and least blended features of OCN$^-$ at around 4.6\,$\mu$m falls in the range observed by Near-Infrared Spectrograph (NIRSpec, \citealt{Jakobsen2022}) mounted on the JWST. Previous laboratory and observational studies have suggested that OCN$^-$ forms in a CO-rich environment and survives in a warmer polar environment (\citealt{Broekhuizen2005}; \citealt{Oberg2011}; \citealt{Boogert2022}). However, observationally we lacked data points for Class 0 objects before the launch of JWST. Thus, only now the formation and evolution of OCN$^-$ can be investigated further through JWST observations. So far a few JWST studies have analyzed the 4.6\,$\mu$m feature of OCN$^-$ in protostellar ices, however, those studies together account for five low- and high-mass protostellar systems (\citealt{Nazari2024_ice}; \citealt{Tyagi2024}). Here we aim to increase the sample size for low-mass objects and study the formation of OCN$^-$.

The only carbon-bearing molecule with at least 6 atoms (i.e., complex organic molecule; COM) that was firmly detected in ices before the launch of JWST, was methanol (CH$_3$OH; \citealt{Gibb2000, Gibb2004}). This was simply due to lack of sensitivity and spectral resolution of previous telescopes to observe species with such low abundances (${\lesssim}10^{-7}$) and low band strengths (a few $10^{-18}$\,cm\,molecule$^{-1}$) even though hints of COMs had been observed (\citealt{Schutte1999}; \citealt{Oberg2011}; \citealt{Terwisscha2021}). Over the past two years JWST has detected several oxygen-bearing COMs for the first time in ices (\citealt{Rocha2024, Chen2024, Rayalacheruvu2025}). Due to the lower abundance of nitrogen in comparison with oxygen, nitrogen-bearing COMs are even more difficult to observe. Among those, the most abundant one in the gas phase is methyl cyanide (CH$_3$CN) and after that few molecules show similar abundances including ethyl cyanide (C$_2$H$_5$CN; e.g., \citealt{Calcutt2018}; \citealt{Yang2021}; \citealt{Nazari2021, Nazari2022}; \citealt{Chahine2022}). Therefore, these may be two of the best candidates when searching for them in ices. In the wavelength range that JWST observes (${\sim}1-28$\,$\mu$m), these two molecules have their strongest transitions at around 4.5\,$\mu$m (\citealt{Moore2010}; \citealt{Rachid2022}). After this, the best transitions are in the ${\sim}7-8$\,$\mu$m regime of the JWST Mid-Infrared Instrument (MIRI, \citealt{Wright2023}). Both CH$_3$CN and C$_2$H$_5$CN have been tentatively detected in protostellar ices of the Investigating Protostellar Accretion (IPA) program by analysis of the 4.5\,$\mu$m feature (\citealt{Nazari2024_ice}). In this work, we aim to expand on that work by increasing the sample size and additionally considering the transitions in the ${\sim}7-8$\,$\mu$m wavelength range. This will also improve our understanding of complex cyanide formation, especially considering the long lasting debate on this topic (\citealt{Huntress1979}; \citealt{Herbst1980}; \citealt{Walsh2014}; \citealt{Garrod2022}; \citealt{Nazari2023}).

Nitrogen is expected to have a refractory and a volatile reservoir. The amount of nitrogen in each component is uncertain, with the dusty refractory component found to be dominant in comet 67P (\citealt{Fray2016}; \citealt{Bardyn2017}; \citealt{Rubin2019}). By updating the band strength of NH$_4^+$, \cite{Slavicinska2025} found that a major component of nitrogen (with abundances of up to ${\sim}20\%$ with respect to water) could be stored in the semi-refractory salts in protostellar systems. Here, we consider putting limits on the volatile N$_2$O and NO ice abundances and investigate whether these molecules could be important nitrogen sinks. N$_2$O ice has been tentatively detected toward other protostellar systems (\citealt{Nazari2024_ice, Rocha2025}), with an ice abundance of just below unity with respect to OCN$^-$ toward Ced 110 IRS4A. Moreover, NO has been recently detected in the gas of protoplanetary disk Oph-IRS 48 (\citealt{Brunken2022}) and model of \cite{Leemker2023} predict that NO and/or N$_2$O may be present in ices.

In this paper, we will study OCN$^-$, CH$_3$CN, C$_2$H$_5$CN, NO, and N$_2$O in ices of the objects covered by the JWST Observations of Young protoStars (JOYS+) program (\citealt{vanDishoeck2025}). The sample covers ${\sim}30$ low- and high-mass protostellar objects (\citealt{Beuther2023}; \citealt{Gieser2023}; \citealt{Francis2024}; \citealt{vangelder2024_SO2}). The low-mass protostars of the sample include both Class 0 and Class I objects. The protostellar systems observed so far by JWST within the JOYS+ program or otherwise show a diversity in their outflow and jet structure (\citealt{Ray2023}; \citealt{Caratti2024}; \citealt{Narang2024}; \citealt{Tychoniec2024}), in addition to their ice and gas-phase spectral features (e.g., \citealt{Federman2024}; \citealt{Neufeld2024}; \citealt{vangelder2024_SO2}; \citealt{Salyk2024}). In the JOYS sample for example, \cite{vanGelder2024} found that although two of the sources (B1-c and L1448-mm) were particularly line-rich (in either emission or absorption), many of them only showed detection of gaseous H$_2$O and CO$_2$ lines. Moreover, \cite{Brunken2024} investigated the ice abundances of CO, CO$_2$, and their isotopologues, in the JOYS+ sample finding that the $^{12}$CO$_2$/$^{13}$CO$_2$ ratios in ices range between ${\sim70-120}$ and the $^{12}$CO/$^{13}$CO are relatively elevated. In terms of COMs, B1-c and IRAS 2A in particular have clearly shown the signature COM absorption features with MIRI (\citealt{Rocha2024}; \citealt{Chen2024}). For this paper, we will focus on a subset of the low-mass sample that has NIRSpec observations (total of 19 sources). 

We first give an overview of the data and the sources in Sect. \ref{sec:obs}. Next, in Sect. \ref{sec:spline} we explain our methods of spline fitting and continuum determination. In Sect. \ref{sec:spec_fit} we describe the spectral fitting and column density measurement. The results for the ratios of CH$_3$CN, C$_2$H$_5$CN, and N$_2$O with respect to OCN$^-$ are presented in Sect. \ref{sec:cyanide_chem}. In Sect. \ref{sec:OCN_chem} we discuss the OCN$^-$/CO and OCN$^-$/CO$_2$ ratios and their significance to OCN$^-$ formation. Finally, we conclude in Sect. \ref{sec:conclusion}.

\section{Observations}
\label{sec:obs}
In this work we use the NIRSpec-IFU (\citealt{Boker2022}) and MIRI-MRS (\citealt{Argyriou2023}) data of the JOYS+ program (PIDs: 1186, 1236, 1290, 1960; PIs: T. P. Greene, M. E. Ressler, and E. F. van Dishoeck). Although the entire JOYS+ sample covers more protostellar systems, only 16 of them have NIRSpec data. As this work primarily uses the NIRSpec data, we focus on those 16 systems. Four of these are binaries, thus resulting in 20 separate Class 0/I objects. From this sample, we remove L1448-IRS1 as it shows a high level of noise and thus conclusions on the targeted molecules in its ices are not particularly meaningful. As a result, 19 objects were analyzed here. The data description and reduction are explained in detail in \cite{vanGelder2024} (MIRI-MRS), \cite{Brunken2024}, and \cite{LeGouellec2025} (NIRSpec), hence we only briefly summarize the observations and analysis here.

The NIRSpec data were taken using either the PRISM or G235H grating at the shorter wavelengths and the G395M or G395H grating at the longer wavelengths. A 4-point dither with the NRSIRS2RAPID readout pattern was used. The data were reduced using additional corrections to the standard JWST pipeline to increase the data quality. More information on the specifics can be found in \cite{Brunken2024} and \cite{LeGouellec2025}. The MIRI-MRS data were mostly taken using a 2-point dither pattern except for B1-c which was observed using a 4-point dither pattern. The data were obtained using the FASTR1 mode. All observations include the four MIRI channels with short, medium, and long gratings fully covering the 4.9-28.6\,$\mu$m range. The data reduction procedure is explained in \cite{vanGelder2024}.

In this work, we use an on-source location (i.e., the peak continuum position at 5.251\,$\mu$m) to extract the spectra. This approach is similar to that used in other JOYS+ papers (\citealt{Rocha2024}; \citealt{Chen2024}; \citealt{Brunken2024}). Therefore, column density ratios of the species considered here and those studied in other JOYS+ papers can be determined referring to the same regions around our objects. The on-source spectral extraction positions in the NIRSpec cubes are given in Table \ref{tab:obs}. There was a slight mismatch between the continuum right ascension (R.A.) and declination (Decl.) at the same wavelength between NIRSpec and MIRI cubes due to astrometry errors of the instruments. This error varies between objects and can be close to zero for some objects but for some others it can be as large as ${\sim}0.5''$. Thus for the MIRI spectra we extracted the spectra from the 5.251\,$\mu$m continuum peak position found in the MIRI cubes rather than using the same R.A. and Decl. found from NIRSpec continuum peak positions. The spectra are extracted in a similar way to \cite{Brunken2024}; using a cone aperture with diameter of 3 times $1.22\lambda/D$ radians, where $D$ is the telescope mirror diameter of 6.5\,m. This method ensures inclusion of a maximum amount of flux without overlap between objects within binaries, although this will be unavoidable at longer MIRI wavelengths. The root mean square (rms) error on the extracted NIRSpec spectra are calculated using the error plane of the data cubes and considering two relatively feature-less regions. In this procedure the error is first turned into optical depth ($\tau$) scale from flux ($f$) scale using

\begin{equation}
    \label{eq:rms}
    \rm{err}_{\tau, \lambda} = \rm{err}_{f, \lambda}/f_{\lambda}  
\end{equation}

\noindent Next, the rms of $\rm{err}_{\tau, \lambda}$ about its mean is reported as the optical depth rms error in Table \ref{tab:obs}.

\section{Band identification, spline, and continuum fitting}
\label{sec:spline}

\begin{figure}
    \centering
    \includegraphics[width=0.9\columnwidth]{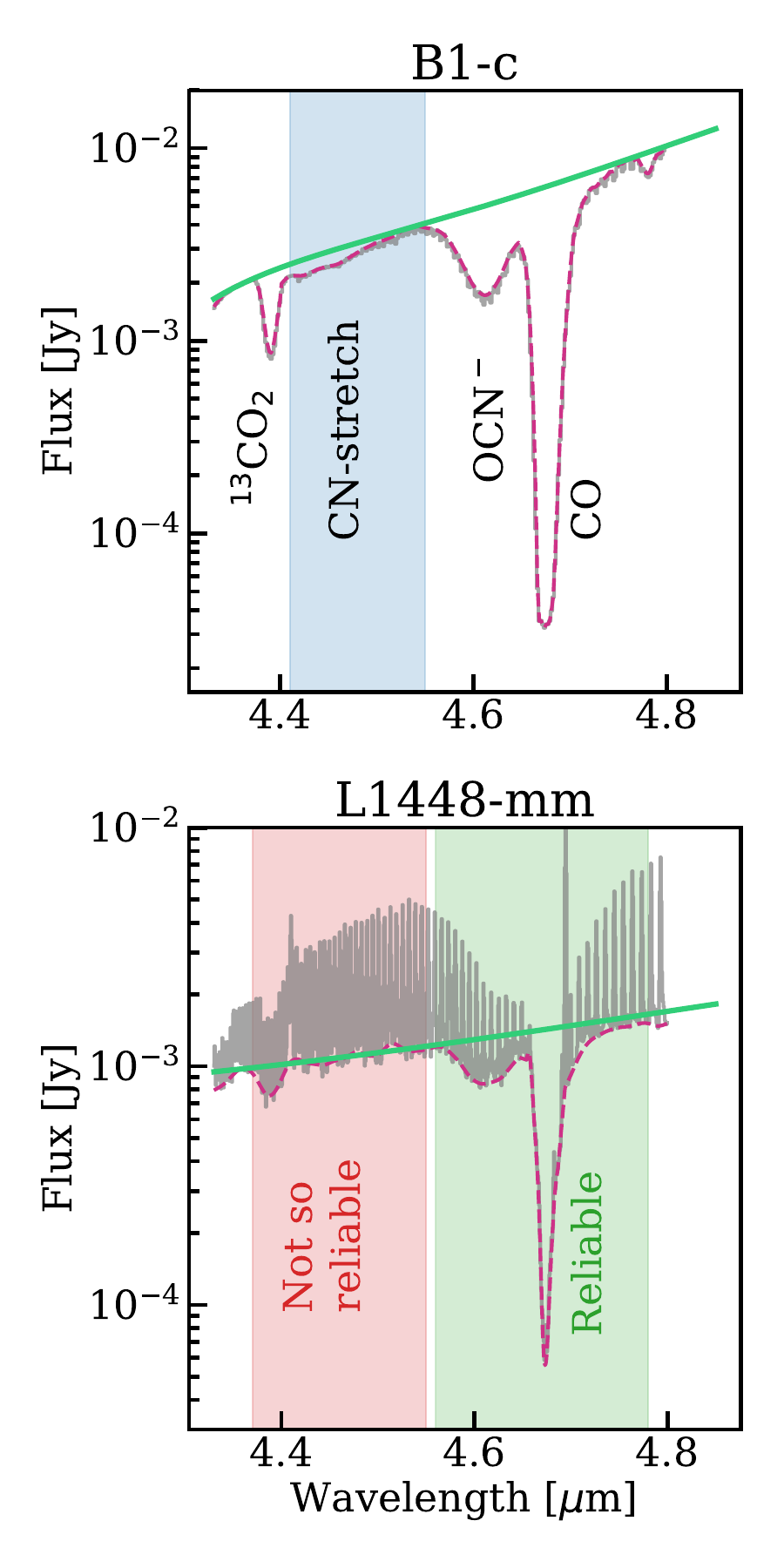}
    \caption{Example of spline fitting for two of our objects. Gray shows the data and dashed magenta line shows the fitted spline to the bottom of the gas-phase emission or top of the gas-phase absorption lines. Most of our objects were too line rich for a reliable measurement of column densities of the weak features at 4.45\,$\mu$m (shaded red area), but the spline fitting was reliable for the stronger features such as OCN$^-$ and CO (shaded green area) even for these sources. An extreme case in gas-phase emission lines is L1448-mm, which is presented in the bottom panel. Both weaker and stronger features of B1-c (top panel) after spline fitting, despite presence of some gas-phase lines, are reliable for further analysis. Green solid line shows the fitted local continuum.}
    \label{fig:spline_text}
\end{figure}

In this paper, we mainly focus on the ${\sim}4.4-4.7$\,$\mu$m region where OCN$^-$ and complex cyanides are expected to have their strongest absorption bands. Unfortunately, CO rotational-vibrational transitions (v=1-0) are also located in this spectral region (\citealt{Federman2024}; \citealt{Rubinstein2024}). They are often detected toward our objects and can affect the ice bands. The other portion of the spectrum (around 5.35\,$\mu$m) used in this work for NO (see Sect. \ref{sec:NO}) also shows gas-phase lines (e.g., from CO and H$_2$O). 

Therefore, we first fitted a baseline (magenta curve in Fig. \ref{fig:spline_text}) to the bottom of the emission lines or top of the absorption lines. This baseline was then used for the ice fitting, although we overlaid our final ice model on the full spectrum with gas-phase lines for consistency check (Sect. \ref{sec:spec_fit}). For this spline fitting, we followed a similar method to \cite{Rubinstein2024} and used the \texttt{pybaselines} package in Python (\citealt{pybaselines}). We used their `penalized spline asymmetric least squares' algorithm via the \texttt{pspline\_asls} function. We altered three main parameters of this function per source to obtain a well-fitted spline to the bottom of the emission lines or the top of the absorption lines. These three parameters were \texttt{lam}, the smoothing parameter, \texttt{num\_knots}, the number of knots for the spline, and \texttt{p}, the penalizing weighting factor which indicates whether to fit the bottom of the gas-phase emission lines or the top of the gas-phase absorption lines. The final fitted baselines to the spectra in the ${\sim}4.4-4.8$\,$\mu$m are shown in Fig. \ref{fig:spline_text} for B1-c (i.e., one of the less extreme objects in showing CO gas-phase lines) and L1448-mm (i.e., one of the most extreme cases in showing bright CO emission lines), while those for the rest of the objects are presented in Fig. \ref{fig:splines_app}.

Next, we fitted a local continuum (green curve in Fig. \ref{fig:spline_text}, $F^{\rm cont}_{\lambda}$) to this region using the polynomial function method of the ENIIGMA fitting tool (\citealt{Rocha2021}). This method is commonly used when analyzing ice bands (e.g., \citealt{Boogert2022}; \citealt{Nazari2024_ice}; \citealt{Slavicinska2024_HDO}). The flux ($F^{\rm data}_{\lambda}$) was then converted to optical depth ($\tau_{\lambda}$) via $-\ln{(F^{\rm data}_{\lambda}/F^{\rm cont}_{\lambda})}$. The fitted continuum in the region between 4.4 and 4.8\,$\mu$m for B1-c and L1448-mm are presented in Fig. \ref{fig:spline_text}, while those for the rest of our objects are shown in Fig. \ref{fig:splines_app}. We particularly took care to fit the continuum as close as possible to the start and end of the OCN$^-$ and CO features to avoid any contamination by the water combination mode at ${\sim}4.5$\,$\mu$m similar to previous works on OCN$^-$ (\citealt{Broekhuizen2005}; \citealt{Boogert2022}).

\section{Spectral fitting and column densities}
\label{sec:spec_fit}
\subsection{OCN$^-$}
\label{sec:OCN}

\begin{table}
\renewcommand{\arraystretch}{1}
    \centering
    \caption{OCN$^-$ ice column densities}
    \label{tab:columns_OCN}
    \begin{tabular}{@{\extracolsep{1.3mm}}*{2}{l}}
          \toprule
          \toprule      
        Source & OCN$^-$ \\
        &  [$\times$ 10$^{16}$ cm$^{-2}$] \\
        \midrule 
 
B1-a-N  &  1.3\\ 
B1-a-S  &  0.8\\ 
B1-b  &  2.7\\ 
B1-c  &  15.4\\ 
EDJ2009183-A  &  0.3\\ 
EDJ2009183-B  &  0.3\\ 
IRAS 2A  &  16.8\\ 
L1448-mm  &  6.5\\ 
L1527  &  7.0\\ 
Per-emb-8  &  3.9\\ 
Per-emb-22  &  8.9\\ 
Per-emb-33  &  8.8\\ 
Per-emb-35  &  12.5\\ 
Per-emb-55-A  &  0.6\\ 
Per-emb-55-B  &  0.2\\ 
S68N  &  5.5\\ 
SMM3  &  5.7\\ 
TMC1-W  &  1.7\\ 
TMC1-E  &  2.9\\

\bottomrule
        \end{tabular}
        \tablefoot{The uncertainties in the column densities are estimated as ${\sim}20\%$ originated from the fits and the data, while the uncertainty in the band strength is negligible (at ${\sim}5\%$, \citealt{Gerakines2025}).}
\end{table}

\begin{figure*}
    \centering
    \includegraphics[width=0.9\textwidth]{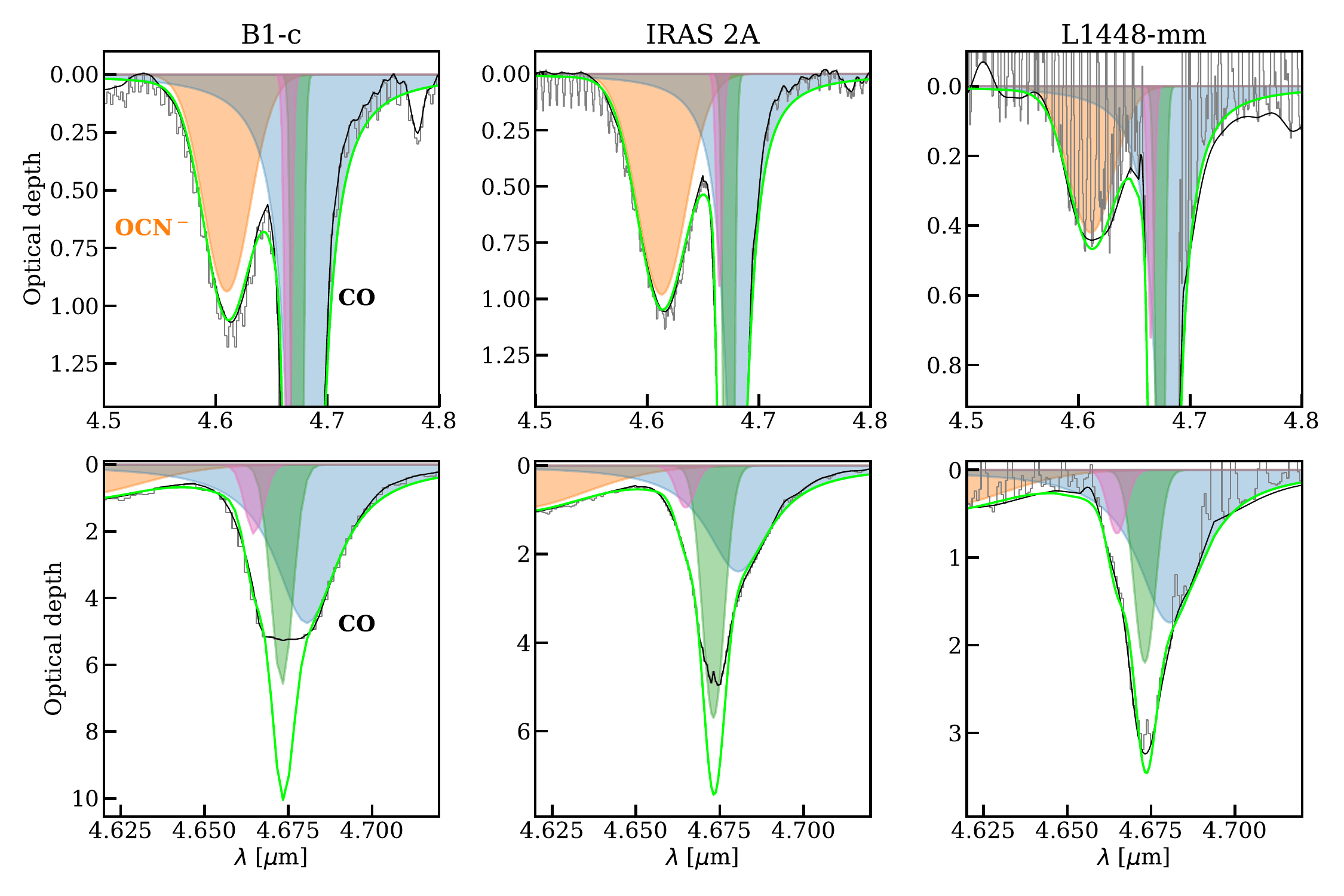}
    \caption{Our fits to the OCN$^-$ and CO absorption features toward three of the studied objects. As explained in the text, OCN$^-$ is fitted simultaneously with CO ice. The orange shaded area shows the fit to the OCN$^-$ feature. Shaded blue, green, and pink areas show the Lorentzian and Gaussian components fitted to the CO feature. The total fit is presented with solid lime green line. The black solid line shows the spline that is fitted to the bottom of the gas-phase emission or top of the gas-phase absorption lines where the data are overplotted in gray. Note that the top and bottom rows have different scales; the top row highlights the OCN$^-$ fits and the bottom row shows the CO fits. For some objects such as B1-c and IRAS 2A, where the CO feature gets saturated close to the noise level, the bottom of the CO feature is ignored when fitting the data and only the wings of the CO band, which are well above the noise level, are fitted. The fits for the other objects considered in this work are given in Figs. \ref{fig:OCN_fits_app} and \ref{fig:CO_fits_app}.}
    \label{fig:OCN_fits}
\end{figure*}

We fit the OCN$^-$ feature at 4.62\,$\mu$m in combination with the adjacent CO ice band following the method used in previous studies using higher spectral resolution data (Fig. \ref{fig:OCN_fits}, \citealt{Pontoppidan2003}; \citealt{Boogert2022}). The CO ice band is expected to be composed of two Gaussian profiles at 2139.9\,cm$^{-1}$ (4.673\,$\mu$m) and 2143.7\,cm$^{-1}$ (4.665\,$\mu$m) with full width at half maximum (FWHM) of 3.5\,cm$^{-1}$ and 3.0\,cm$^{-1}$, and one Lorentzian profile at 2136.5\,cm$^{-1}$ (4.681\,$\mu$m) with FWHM of 10.6\,cm$^{-1}$. These three components indicate CO in different ice environments. The broad most redshifted Lorentzian profile is expected to be CO mixed with polar species such as H$_2$O and CH$_3$OH and is thus referred to as CO$_{\rm polar}$ component. The Gaussian profile in the middle is expected to be CO in an apolar environment and thus is referred to as CO$_{\rm apolar}$. The minor shortest-wavelength Gaussian component is likely CO mixed with CO$_2$ ice and often referred to as CO$_{\rm blue}$ (\citealt{Tielens1991}; \citealt{Boogert2002}; \citealt{Pontoppidan2003}). We opted to include these three components as opposed to a single Gaussian in our fitting. This choice was made after a few tests on the higher spectra resolution data ($R{\sim}10000$) of \cite{Pontoppidan2003}. We first convolved the data used in that study for a few objects to the NIRSpec spectral resolution ($R{\sim}2700$). Next, we applied our fitting procedure to both the data and its convolved version. We found similar fits for the three components to well within $10\%$ which led us to follow the same procedure as \cite{Pontoppidan2003}.

The feature at ${\sim}4.6$\,$\mu$m was decomposed into two Gaussian profiles (at 2165.7\,cm$^{-1}$, 4.617\,$\mu$m, with FWHM of 26\,cm$^{-1}$ and at 2175.4\,cm$^{-1}$, 4.597\,$\mu$m, with FWHM of 15\,cm$^{-1}$) by \cite{Broekhuizen2005} using the higher spectral resolution VLT data. The major component at 2165.7\,cm$^{-1}$ is often firmly associated with OCN$^-$ (\citealt{Broekhuizen2005}; \citealt{Oberg2011}). However, given the lower spectral resolution of JWST, the complexities with the gas-phase lines and their removal from the ${\sim}4.6$\,$\mu$m feature (Sect. \ref{sec:spline}), we only consider one Gaussian to fit the data. We note that these issues are less important for the CO band as there are smaller number of weaker CO ro-vibrational lines on top of the narrower CO ice band. We leave the single Gaussian parameters for OCN$^-$ to vary freely. 

In total, we fit the 4.5-4.75\,$\mu$m region with four profiles (one for OCN$^-$ and three for CO) simultaneously using the \texttt{curve\_fit} function in Python. We fit the spectra obtained from the spline fitting (i.e., those with gas-phase lines removed). The final fits are presented in Figs. \ref{fig:OCN_fits} and \ref{fig:OCN_fits_app} (see \ref{fig:CO_fits_app} for the CO fits) overlaid with the data including gas-phase lines to emphasize the good match. These figures show that OCN$^-$ is detected toward all our objects and the final model (lime green line) matches well with the data. They also show that OCN$^-$ feature is strong enough to be fitted regardless of the uncertainty associated with the spline fitting. 

The comparison of the FWHM and peak position from our single Gaussian profiles with the components considered in \cite{Broekhuizen2005} is presented in Fig. \ref{fig:OCN_compare}. For ${\sim}90$\% of our objects both the peak position and FWHM fall in between the range considered by \cite{Broekhuizen2005} for their two components. However, the FWHM of our Gaussian profiles for two objects (EDJ2009183-A and Per-emb-55-B) are smaller than than 15\,cm$^{-1}$ (i.e., FWHM for the minor component considered in \citealt{Broekhuizen2005}). This is likely due to the weakness of the OCN$^-$ feature and strong gas-phase emission lines for these two objects in our sample (Fig. \ref{fig:OCN_fits_app}). 

After the fitting, we calculated the column density of OCN$^-$ and CO components using

\begin{equation}
    N = \int{\tau_{\tilde\nu} d\tilde\nu}/A,
    \label{eq:column}
\end{equation}

\noindent where $\tau_{\tilde\nu}$ is the optical depth at wavenumber $\tilde\nu$ and $A$ is the band strength. We take the band strengths of CO and OCN$^{-}$ from new laboratory measurements of \cite{Gerakines2023} and \cite{Gerakines2025} as $1.1 \times 10^{-17}$\,cm\,molecule$^{-1}$ and $1.5 \times 10^{-16}$\,cm\,molecule$^{-1}$ (Table \ref{tab:band_strengths}). It is worth noting that \cite{Pontoppidan2003} introduced an additional factor of 0.71 to be multiplied by $A_{\rm CO}$ when measuring the column density of the middle CO$_{\rm apolar}$ from Eq. \eqref{eq:column}. This was to take into account grain shape effects in the Rayleigh limit, in which case the band strength of the middle CO$_{\rm apolar}$ component is expected to decrease by 29\%. However, in our calculations the same $A$ of $1.1 \times 10^{-17}$\,cm\,molecule$^{-1}$ has been assumed for all three components of CO, and we do not consider this additional factor. In other words, including that factor would increase our column densities of CO$_{\rm apolar}$ by 1.41 (i.e., 1/0.71). Nevertheless, either approach is an assumption and a more detailed study of grain shape and size effects on column densities is needed. Finally, for some objects (B1-b, B1-c, IRAS 2A, L1527, Per-emb-8, Per-emb-22, and Per-emb-33) the bottom of the CO band approaches the noise level in flux scale and thus is not reliable. For those, we only fit the wings of the CO feature to estimate the true optical depth of CO.

We note that the total CO ice column densities for our sample have been measured by \cite{Brunken2024}. We opted to re-measure those but now considering OCN$^-$ contribution to the fits simultaneously because \cite{Brunken2024} simply integrated the data over the CO absorption feature (${\sim}4.65-4.72\,\mu$m). The band strength used for CO in our work is $1.1 \times 10^{-17}$\,cm\,molecule$^{-1}$ from new laboratory experiments (\citealt{Gerakines2023}), while the one used in \cite{Brunken2024} from older studies (\citealt{Gerakines1995}; \citealt{Bouilloud2015}) was $1.4 \times 10^{-17}$\,cm\,molecule$^{-1}$ (i.e., a factor of 1.27 larger). We report the column densities of OCN$^-$ in Table \ref{tab:columns_OCN} and report the column densities of different CO components with their sums in Table \ref{tab:columns_CO}. The uncertainty in the column densities is of the order of 20\%, considering the uncertainty in the fits, data, and continuum subtraction (see the discussion in Appendix A of \citealt{Brunken2024}). The band strengths for CO and OCN$^-$ from the new laboratory work have uncertainties that are negligible (${\sim}5\%$, \citealt{Gerakines2023, Gerakines2025}). We note that the band strengths would change for ices in different ice matrices but are not expected to change by more than ${\sim}$10-20\%. Within uncertainties, our OCN$^-$ column densities for B1-c and IRAS 2A are consistent with those found from the OCN$^-$ transition at 7.6\,$\mu$m (\citealt{Chen2024}; \citealt{Rocha2024}), providing a confirmation for the reliability of both methods. Our total CO column densities agree with those of \cite{Brunken2024} after correction for the different band strengths used, within ${\sim}15-20\%$ for about half of our objects (i.e., those with minimal contribution from OCN$^-$ to CO or vice versa). However, for the other half, the difference is larger and can get up to a factor of ${\sim}1.5$. These objects are mostly those with a saturated CO band, where we fit the wings of the CO feature and thus estimate larger CO column densities than those obtained by direct integration of the observed feature.

\begin{figure*}
    \centering
    \includegraphics[width=\textwidth]{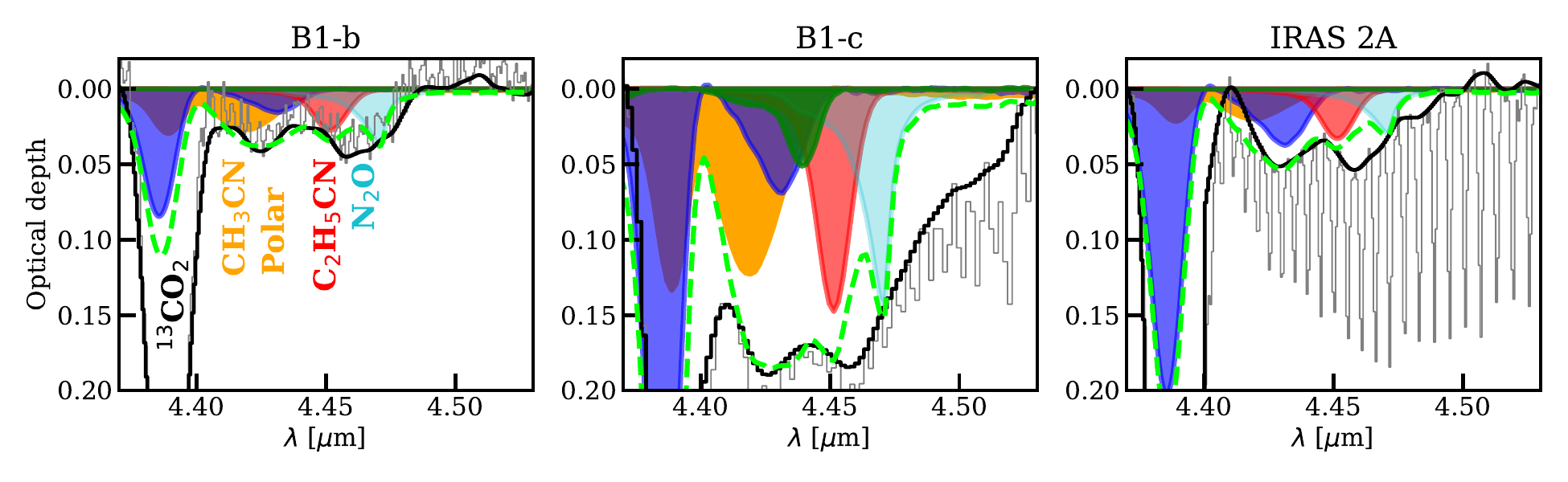}
    \caption{Final fits to the cyanide stretch for objects showing tentative detections of complex cyanides and N$_2$O. Black line shows the spline fitted to the bottom of the gas-phase emission or top of the gas-phase absorption lines where gray shows the data. Orange and blue show the ice mixtures of CH$_3$CN:H$_2$O:CO$_2$  (1:5:2, 50\,K) and CH$_3$CN:CO$_2$ (1:10, 15\,K; polar ices). Green shows the CH$_3$CN:CO ice mixture (1:10, 80\,K; apolar ices), red shows the C$_2$H$_5$CN component (50\,K) and cyan N$_2$O (70\,K). The total fit is shown as a lime green dashed line.}
    \label{fig:CH3CN_fits}
\end{figure*}

\subsection{Complex cyanides and N$_2$O}
\label{sec:cyanides_N2O}
\subsubsection{NIRSpec range}
\label{sec:NIRSPEC_tentative}
The absorption bands of complex cyanides and N$_2$O are relatively weak. The detection of these weaker features are less robust especially for the more line-rich objects. Therefore, we considered each object individually and for this section we only focus on the subset of sources where the CN-stretching absorption feature is detected and does not seem like an artifact when the spline fitted to the data is compared to the data themselves. We fit the CN-stretching band by eye using the same components as \cite{Nazari2024_ice}, where the minimum number of components from the available laboratory data was found to fit the CN-stretching band. These components are CH$_3$CN ice mixtures, C$_2$H$_5$CN, and N$_2$O. The laboratory spectra for these molecules are taken from \cite{Rachid2022}, \cite{Moore2010}, and \cite{Gerakines2020}, respectively. It is worth noting that water has a broad band due to its combination mode at 4.5\,$\mu$m. However, because of the way that the local continuum is fitted (i.e., as close to the ice features as possible, Fig. \ref{fig:spline_text}) the water combination mode is already subtracted before any further fitting to the ice bands. We tested that this is indeed the case for B1-c. 

The final fits are shown in Fig. \ref{fig:CH3CN_fits}. We emphasize that these fits are not unique, but scaling these components differently to fit the data will affect the column densities by less than a factor of ${\sim}2$. Moreover, we only claim a tentative detection of CH$_3$CN, C$_2$H$_5$CN, and N$_2$O, considering the uncertainties that are attributed to the final fits. Another reason to stay conservative is that although the current fits presented in Fig. \ref{fig:CH3CN_fits} are based on the minimum number of components required from the available laboratory data, we cannot rule out the possibility of laboratory data that are missing and could fit this region with smaller number of components. More specifically, Fig. \ref{fig:CH3CN_fits} consistently hints to an absorption feature in between the C$_2$H$_5$CN and N$_2$O peaks at ${\sim}4.46$\,$\mu$m with an unknown identity. Therefore, that feature may be explained by one component with a smaller contribution from C$_2$H$_5$CN and N$_2$O. Potentially, this component could be an ice mixture of C$_2$H$_5$CN or N$_2$O that could easily be shifted from the pure spectra shown here. However, to the best of our knowledge the laboratory spectra of these mixed ices are not yet available. We use Eq. \eqref{eq:column} to find the column densities of CH$_3$CN, C$_2$H$_5$CN, and N$_2$O which are given in Table \ref{tab:cyanide_columns} and their band strengths in Table \ref{tab:band_strengths}. The uncertainty in the column density is dominated by the continuum determination and that of the band strength. We estimate an uncertainty of around 50\% on our column densities of CH$_3$CN, C$_2$H$_5$CN, and N$_2$O.

\begin{table}
\renewcommand{\arraystretch}{1.3}
    \caption{Ice column densities of CH$_3$CN, C$_2$H$_5$CN, and N$_2$O.}
    \label{tab:cyanide_columns}
    \resizebox{\columnwidth}{!}{\begin{tabular}{@{\extracolsep{1mm}}*{4}{l}}
          \toprule
          \toprule      
        Sources & $N_{\rm CH_3CN}\, (\rm cm^{-2})$ &  $N_{\rm C_2H_5CN}\, (\rm cm^{-2})$ &  $N_{\rm N_2O}\, (\rm cm^{-2})$\\
        \midrule 

B1-b  &  2.4 $\times$ 10$^{17}$  &  8.9 $\times$ 10$^{16}$  &  5.6 $\times$ 10$^{15}$  \\
B1-c  &  1.4 $\times$ 10$^{18}$  &  4.8 $\times$ 10$^{17}$  &  2.3 $\times$ 10$^{16}$  \\
IRAS 2A  &  3.8 $\times$ 10$^{17}$  &  1.1 $\times$ 10$^{17}$  &  4.5 $\times$ 10$^{15}$  \\
\bottomrule
        \end{tabular}}
        \tablefoot{The uncertainties for these species are estimated as 50\%, dominated by the continuum fitting and the uncertainty in the band strength.}
\end{table}

\subsubsection{MIRI range}

After the ${\sim}4.5$\,$\mu$m feature, the second best wavelength range to search for CH$_3$CN, C$_2$H$_5$CN, and N$_2$O is ${\sim}6.8-8$\,$\mu$m which can be used for consistency checks. The ${\sim}6.8-8$\,$\mu$m range is also where many other complex organic molecules have a strong transition (\citealt{Yang2022}; \citealt{Rocha2024}; \citealt{Chen2024}). Therefore, initial modeling of that regime with other molecules is needed (\citealt{Rayalacheruvu2025}). Thus, here, we only consider the two objects in our sample for which this modeling has been done, namely B1-c (\citealt{Chen2024}) and IRAS 2A (\citealt{Rocha2024}). Figure \ref{fig:MIRI_fit} presents the total fit on top of the continuum- and silicate-subtracted spectra from these two papers, where we have added CH$_3$CN, C$_2$H$_5$CN, N$_2$O lab spectra scaled based on our fits to the ${\sim}4.5$\,$\mu$m regime. For IRAS 2A (top panel of Fig. \ref{fig:MIRI_fit}), most new features are small enough such that the total fit remains relatively consistent with the data. However, for B1-c our column densities from the 4.5\,$\mu$m band are not entirely consistent with the fit at ${\sim}6.8-8$\,$\mu$m (bottom panel of Fig. \ref{fig:MIRI_fit}). This may be related to CH$_3$CN, C$_2$H$_5$CN, and N$_2$O not being responsible for the entire ${\sim}4.5$\,$\mu$m absorption feature. However, N$_2$O has also been tentatively detected toward the Ced 110 IRS4A protostellar system at 7.75\,$\mu$m (\citealt{Rocha2025}). Alternatively, (local) continuum- and silicate subtraction uncertainties in the ${\sim}6.8-8$\,$\mu$m region could play a role. As the cyanide features are relatively weak compared to other COMs studied by \cite{Chen2024} and \cite{Rocha2024}, we refrain from claiming a detection for our molecules even after considering the MIRI range. 

\begin{figure*}
    \centering
    \includegraphics[width=\textwidth]{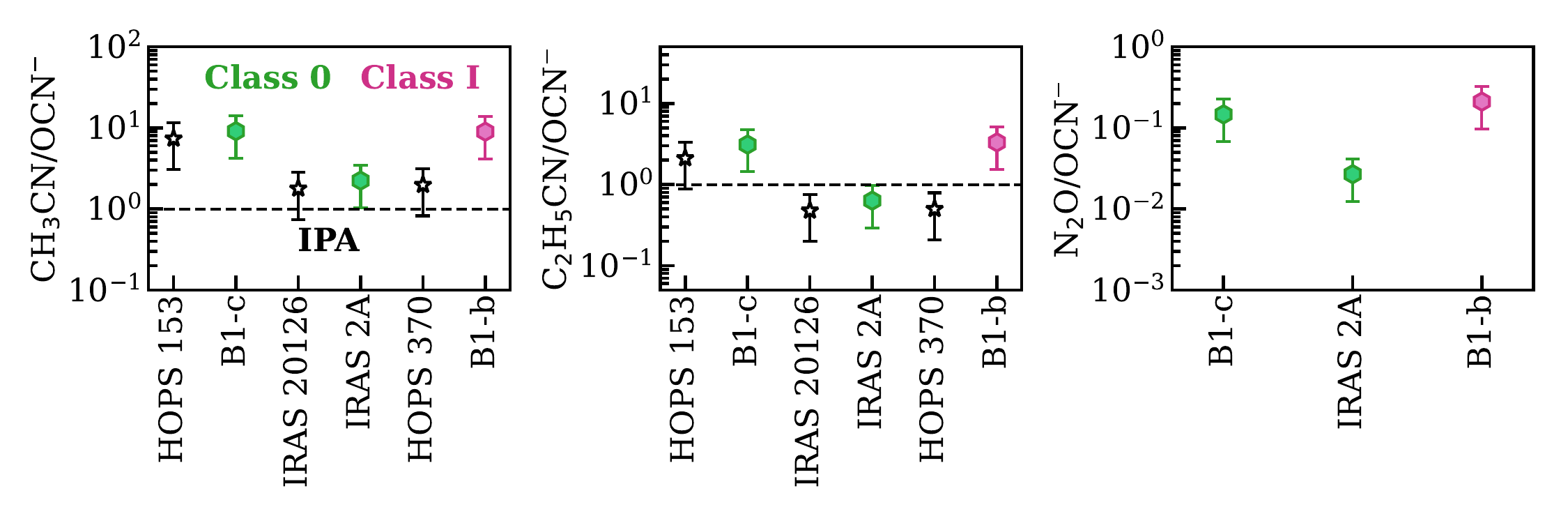}
    \caption{Ice column density ratios of nitrogen-bearing molecules considered here. Green represents Class 0 objects and pink Class I objects. Black data points are results of the IPA JWST program, updated for the OCN$^-$ band strength used here (i.e., ${\sim}15\%$ higher, \citealt{Nazari2024_ice}). The objects are ordered by their bolometric temperature ($T_{\rm bol}$) from left to right.}
    \label{fig:ratios_CNs}
\end{figure*}

\subsection{Silicate and H$_2$O}
\label{sec:silicate_H2O}

We aim to consider the ratios of our OCN$^-$ measurements with respect to H$_2$O as well as the relation between the column densities and $A_{\rm V}$ to estimate the formation timeline of OCN$^-$. Therefore, we measure $A_{\rm V}$ by fitting the ${\sim}9.8$\,$\mu$m silicate feature and multiplying $\tau_{\rm peak}$ of that feature by 28.6 (based on dense clouds) and 18.5 (based on diffuse clouds) for Class 0 and I objects (\citealt{Boogert2013}; \citealt{Madden2022}; \citealt{vanDishoeck2025}). The reason for this difference is that \cite{vanDishoeck2025} found unreasonably high accretion rates if the same conversion of 28.6 was applied to Class I objects. Therefore, they used the relation found from diffuse clouds for those (see their Appendix G) which we follow. The silicate at ${\sim}9.8$\,$\mu$m and water absorption at 13.6\,$\mu$m are fitted together following the same method as \cite{Rocha2024} and \cite{Chen2024}. Particularly, we first fit a global polynomial continuum to the entire MIRI spectrum of the objects and after subtraction of this continuum, we fit a mixture of synthetic silicates (consisting of small and large pyroxene and olivine grains) to the 9.8\,$\mu$m band. Next the water at 13.6\,$\mu$m is fitted. Normally, this method results in reasonable total fits but in some cases iteration is needed between water and silicate fitting to ensure a good fit. This fitting process is done by hand, using an interactive code that is introduced and explained in detail in Chen et al. (in preparation). Our method does not consider the contribution from carbon and thus our $A_{\rm V}$ values may be lower limits. Nevertheless, this effect is not expected to change the values by more than ${\sim}20\%$.

The water band for most objects could be fitted with a cold (15\,K) and warm (150\,K) component using the laboratory spectra of \cite{Slavicinska2025}. The band strengths for column density measurement of water are given in Table \ref{tab:band_strengths}. We also note that our fits are not unique, and thus it is important to consider the final values as approximate. To better constrain the range of possible values, we vary the continuum and silicate fitting to obtain multiple reasonable fits and we report the minimum and maximum possible $A_{\rm V}$ considering those different fits. The final $A_{\rm V}$ and water column densities represent the middle of that range per object, and are generally in good agreement with those found by Chen et al. (in preparation) within ${\sim}50\%$. Table \ref{tab:columns_water} presents the results of this fitting for $A_{\rm V}$ and water.

\subsection{NO}
\label{sec:NO}

Another molecule that we considered in this work is NO. This is stimulated by the detection of NO in the gas phase toward the Class II object, Oph-IRS 48, at the location of the ice trap in the disk where ices are thought to have thermally sublimated (\citealt{Brunken2022}; \citealt{Booth2024}). Moreover, N$_2$O is tentatively detected in our ices (Sect. \ref{sec:cyanides_N2O}) and there might be a chemical connection between N$_2$O and NO from chemical models of \cite{Leemker2023}. As shown in laboratory work by \cite{Fulvio2019}, NO has a peak position at 1869\,cm$^{-1}$ (or 5.35\,$\mu$m) which falls in the MIRI range. We thus searched for NO in all of our objects. After spline fitting to remove the gas-phase lines and a local continuum subtraction, five objects showed a tentative absorption feature at 5.35\,$\mu$m as presented in Fig. \ref{fig:NO} (their MIRI optical depth rms error ranged between ${\sim}5\times 10^{-4} - 5\times 10^{-3}$). Assuming that the entire absorption feature at 5.35\,$\mu$m is due to NO, we find upper limits of ${\sim}3\times 10^{16} -10^{17}$\,cm$^{-2}$ for the five objects. In our by-eye fitting, we fixed the FWHM of NO to 20\,cm$^{-1}$ based on the NO ice feature that were produced from irradiation of NO$_2$:N$_2$O$_4$ ice in \cite{Fulvio2019} (see their Figs. 6 and 7). The column densities were calculated using Eq. \eqref{eq:column} and assuming a band strength of $6.8 \times 10^{-18}$\,cm\,molecule$^{-1}$ (\citealt{Stirling1994}; \citealt{Jamieson2005}; \citealt{Fulvio2019}). Our calculated NO/OCN$^-$ upper limits are ${\sim}0.2-3$. Considering the same objects, the NO/OCN$^-$ upper limits are a factor of ${\sim}1-10$ higher than the N$_2$O/OCN$^-$ ratios and a factor of ${\sim}2-100$ lower than those for CH$_3$CN/OCN$^-$ (Fig. \ref{fig:ratios_CNs}). Thus this indicates that neither N$_2$O nor NO store the bulk of nitrogen in ices.

We can also compare our values with what \cite{Leemker2023} predicted for NO in ices to match the gas-phase observations of NO toward Oph-IRS 48. They found that an abundance of ${\sim}5\times 10^{-6}-10^{-5}$ with respect to hydrogen ($N_{\rm H} = 1.6 \times 10^{22}$\,cm$^{-2}$) was needed in the ices. Converting our measured visual extinctions ($A_{\rm V}$, see Table \ref{tab:columns_water} and Sect. \ref{sec:silicate_H2O}) to $N_{\rm H}$ using $1.37 \times 10^{21} A_{\rm V}$ (\citealt{Evans2009}), we find $N_{\rm NO}$/$N_{\rm H}$ abundances of ${\sim}2\times 10^{-7}-10^{-6}$ which are around one order of magnitude smaller than those required for NO in ices to reproduce the observations of gas-phase NO toward Oph-IRS 48 (\citealt{Leemker2023}). This might further point to the gas-phase NO being a secondary product of other ice species such as N$_2$O as discussed in \cite{Leemker2023}. No explicit N$_2$O ice chemical network was included in their models. However, the detected N$_2$O in our objects have abundances of ${\sim}2\times 10^{-8}-10^{-7}$ which are in good agreement with the NO gas-phase abundances toward Oph-IRS 48, making N$_2$O a likely parent molecule of NO after desorbing from the ices.

\subsection{Ice ratios}
\label{sec:cyanide_chem}

Figure \ref{fig:ratios_CNs} presents the ice column density ratios of CH$_3$CN, C$_2$H$_5$CN, and N$_2$O with respect to OCN$^{-}$. The results of the cyanides ice analyses from the IPA program are also plotted (\citealt{Nazari2024_ice}). Generally, there is good agreement between column density ratios of different sources. The column density ratios range within ${\sim}1$ order of magnitude for various objects. These constant column density ratios point to similar ice environment of CH$_3$CN, C$_2$H$_5$CN, and N$_2$O with OCN$^-$ and may indicate that the complex cyanides form along with OCN$^-$. In Sect. \ref{sec:OCN_chem} we discuss the formation of OCN$^-$ in detail. \cite{Nazari2024_ice} compared the ice ratios in the IPA sample with the gas-phase data. They found that the ratios were roughly similar between the gas and ice where the differences were associated with various physical and chemical effects. Given the good agreement of our ice results with those from the IPA data, similar conclusions hold.

The ratios of CH$_3$CN/OCN$^-$ are between ${\sim}1-10$ which are the largest compared with ratios of C$_2$H$_5$CN and N$_2$O with respect to OCN$^-$. The C$_2$H$_5$CN/OCN$^-$ ratios vary between ${\sim}0.2-2$ and N$_2$O/OCN$^-$ ratios vary between ${\sim}0.01-0.1$. Out of these three molecules, CH$_3$CN could be an important nitrogen reservoir given that it could have an ice column density as large as 10 times the amount of OCN$^-$, pushing it close to the abundances observed for NH$_3$ in low-mass protostellar ices (\citealt{Bottinelli2010}; \citealt{Oberg2011}). Nevertheless, a somewhat larger fraction of nitrogen is likely in salty refractories (NH$_4^+$), as it was found by \cite{Slavicinska2025}.

The few objects in the IPA program that showed tentative detection of CH$_3$CN were also those with evidence for ice thermal processing seen as a double peak feature in the $^{13}$CO$_2$ band (\citealt{Brunken2024IPA}) and those with the highest luminosities. This led \cite{Nazari2024_ice} to conclude that there may be a tentative evidence for enhancement of CH$_3$CN in thermally processed and warmer ices. However, with the larger sample in this work, we do not find this trend to hold anymore. Inspecting the $^{13}$CO$_2$ bands in our sample as presented in \cite{Brunken2024} (also see Fig. \ref{fig:spline_text} and \ref{fig:splines_app}), we find objects that do show a (tentative) CH$_3$CN detection but do not seem to have thermally processed ices (e.g., B1-c) or vice versa (e.g., L1527 and Per-emb-35). Therefore, there is no strong evidence for enhancement of CH$_3$CN in warmer ices from this larger sample.

\section{OCN$^-$ formation}
\label{sec:OCN_chem}

\begin{figure}
    \centering
    \includegraphics[width=0.8\columnwidth]{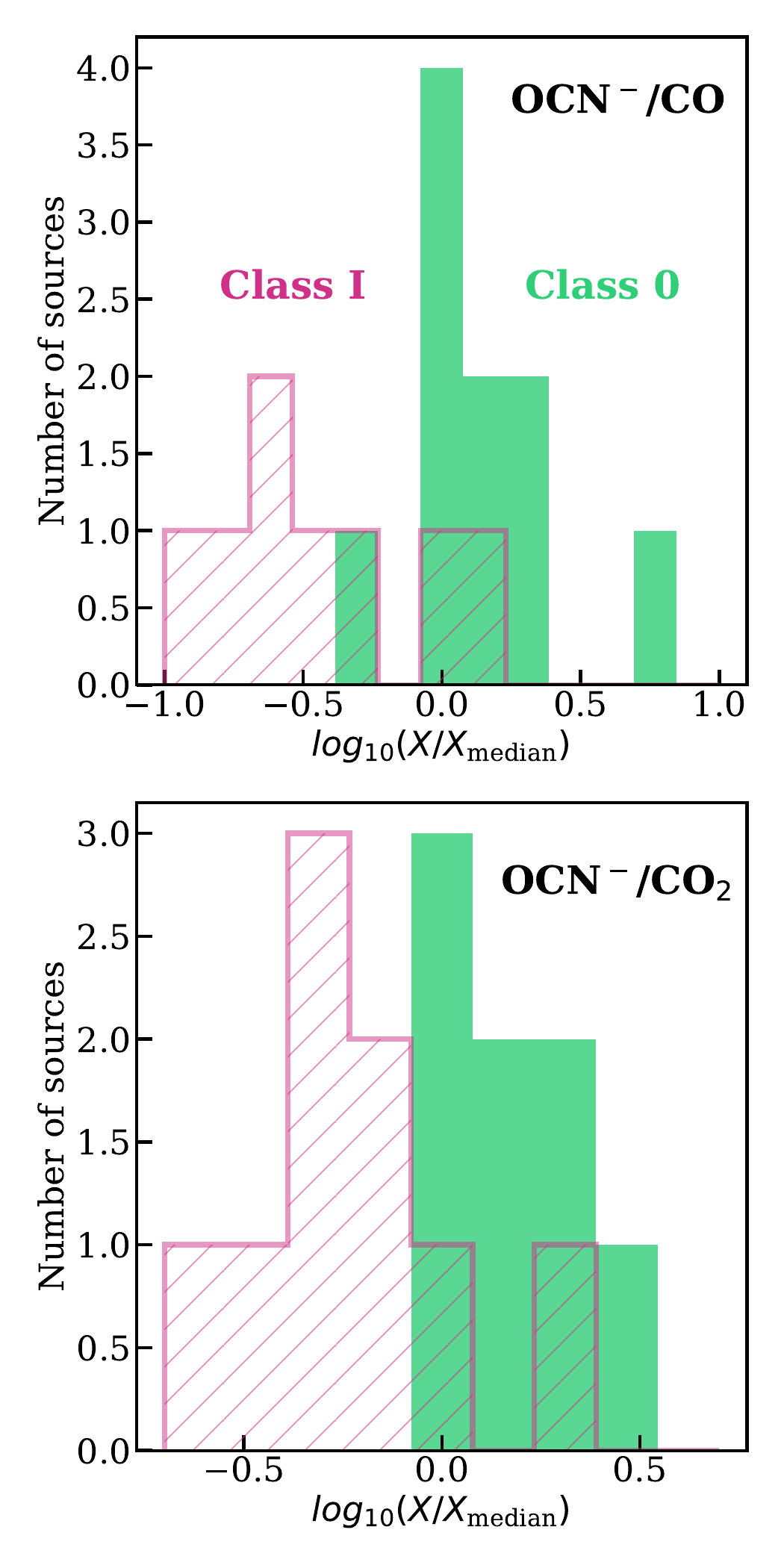}
    \caption{Histogram of Class 0 (smooth green) and I (hatched pink) objects distributed based on $log_{10}(X/X_{\rm median})$, where $X$ is OCN$^-$/CO in the top panel and OCN$^-$/CO$_2$ in the bottom panel. The CO$_2$ column densities are taken from \cite{Brunken2024}. The Class 0/I objects have been assumed as Class 0 in this figure. There is a separation between Class 0 and I objects in both OCN$^-$/CO and OCN$^-$/CO$_2$.}
    \label{fig:OCN_CO_CO2}
\end{figure}

Previous studies using ground-based telescopes could either study ices of high-mass protostellar systems or bright low-mass systems which were mainly Class I objects (e.g., see review by \citealt{Boogert2015}). With JWST we can, for the first time, also study OCN$^-$ in Class 0 sources. Figure \ref{fig:OCN_CO_CO2} presents histograms of our objects divided based on their OCN$^-$/CO and OCN$^-$/CO$_2$ ice ratios. This figure interestingly shows a divide between Class 0 and I objects. The Class 0 and those on the border of Class 0/I in general have larger OCN$^-$/CO and OCN$^-$/CO$_2$ than late Class Is. This might be related to contribution from the ices of the disk for the more evolved objects. However, to investigate this further detailed radiative transfer models including ices (i.e., similar to those of \citealt{Sturm2023} and \citealt{Bergner2024} for Class IIs) that consider the evolution from Class 0 to Class I are required which is outside the scope of this work. 

\subsection{Thermal processing}
\label{sec:thermal}

\begin{figure}
    \centering
    \includegraphics[width=\columnwidth]{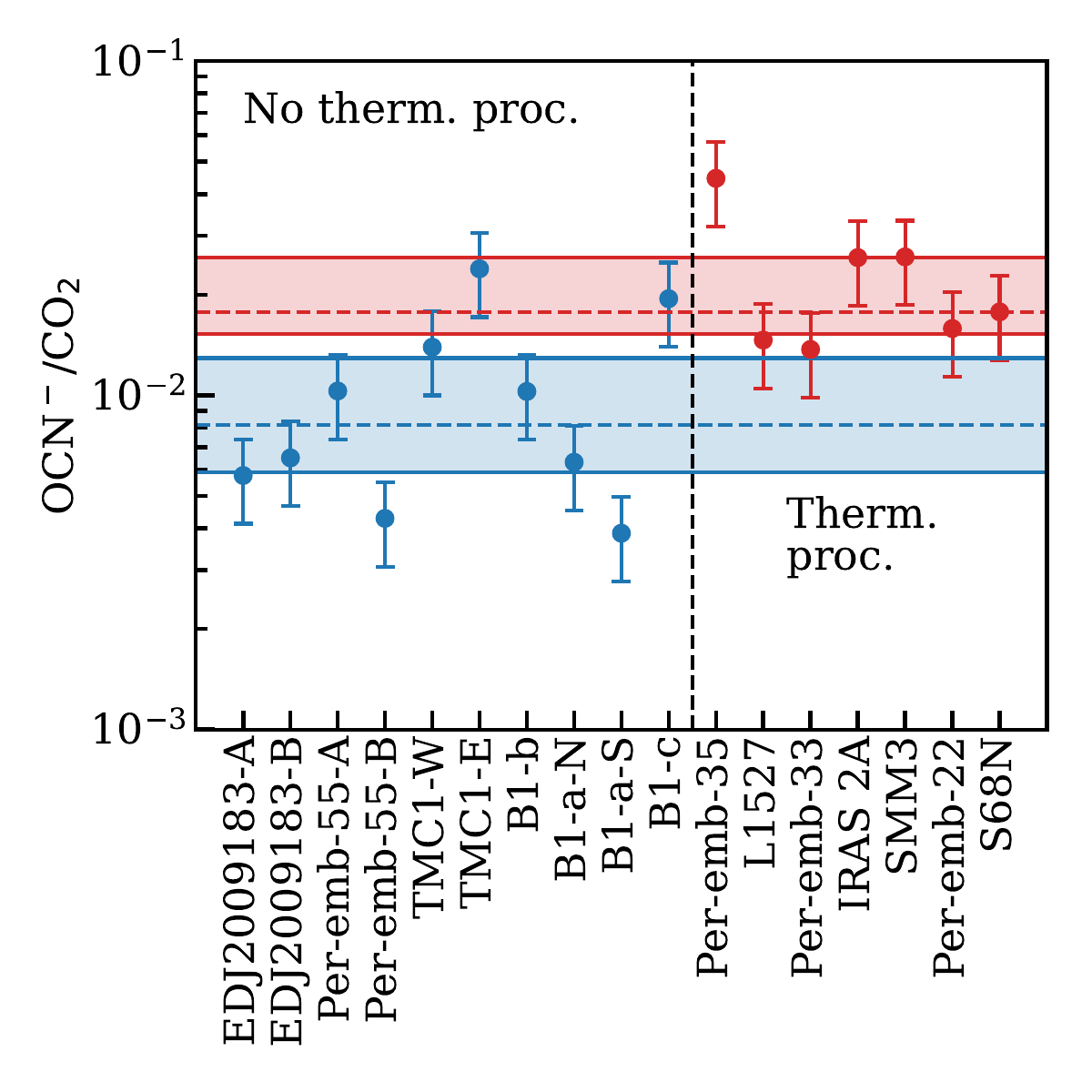}
    \caption{Effect of heating on OCN$^-$/CO$_2$ ratio. The objects are divided into two groups based on the shape of the CO$_2$ band. Those with double-peaked CO$_2$ are shown in red (thermally processed ices) while those with a single-peaked CO$_2$ are shown in blue (no thermal processing). The dashed lines show the median of each group and the shaded areas show the range where the data falls in between the upper and lower quartiles. Ice heating can enhance OCN$^-$ formation by only a factor of ${\sim}2-3$.}
    \label{fig:thermal}
\end{figure}

To assess the effect of heating on formation of OCN$^-$, Fig. \ref{fig:thermal} presents the column density ratios of OCN$^-$ with respect to CO$_2$. The sources are divided in two groups based on the shape of their 15.2\,$\mu$m CO$_2$ absorption band. The objects with a double-peak morphology are shown in red and those with a single-peak morphology are shown in blue. Those with a double peak CO$_2$ feature are thought to indicate thermally processed ices (\citealt{Pontoppidan2008}; \citealt{Brunken2025}). Figure \ref{fig:thermal} shows that objects with thermally processed ices (double-peak CO$_2$ morphology) statistically have a factor of ${\sim}2-3$ larger OCN$^-$/CO$_2$ than those ices without thermal processing. The same general trend holds for OCN$^-$/H$_2$O ratios toward our objects even though that is less certain considering the complexities in water column density measurements (Sect. \ref{sec:silicate_H2O}). Therefore,
the enhanced OCN$^-$/CO$_2$ ratios are likely not caused by CO$_2$ sublimation for the objects with more thermally processed ices.

This factor of ${\sim}2-3$ enhancement is in excellent agreement with the results from \cite{Broekhuizen2004} laboratory experiments. They found that OCN$^-$ can form more efficiently at higher temperatures where the difference between OCN$^-$ abundance at 20-40\,K and 120-140\,K was a factor of ${\sim}2-3$. They found that UV processing can also enhance the formation of OCN$^-$ but that this required a much larger UV fluence in the laboratory than experienced by grains in molecular clouds. Based on those results and Fig. \ref{fig:thermal}, we conclude that simply detecting OCN$^-$ is not a diagnostic of energetic processing of ices and the effect of heating on OCN$^-$ enhancement is only a factor of ${\sim}2-3$ across Class 0 and I objects.

\subsection{Formation timeline with respect to CO, CO$_2$, and H$_2$O}
\label{sec:timeline}

It has been suggested that OCN$^-$ forms through the acid-base reaction of HNCO and NH$_3$, while HNCO itself can form via reactions of NH and NH$_2$ with CO ice (\citealt{Fedoseev2016}; \citealt{Ligterink2018}). Because of the high sublimation temperature of salts (${\sim}200$\,K in Temperature Programmed Desorption experiments, \citealt{Ligterink2018}) OCN$^-$ can survive the later ice processing in a polar environment (\citealt{Broekhuizen2005}; \citealt{Oberg2011}; \citealt{Boogert2022}). In this section we investigate this further by considering the relation between OCN$^-$ column densities and the visual extinction ($A_{\rm V}$) of our objects. In \cite{Boogert2015} the x-intercepts of column densities of H$_2$O, CO$_2$, and CO in molecular clouds as a function of $A_{\rm V}$ are suggested to indicate the formation order of these molecules. Here, we attempt to examine the same relation for OCN$^-$, H$_2$O, and CO$_2$ in our objects, keeping in mind that in our case heating might affect the linearity of the relation and that a large extrapolation is needed to find the x-intercepts due to the large visual extinction of protostellar objects.

\begin{figure}
    \centering
    \includegraphics[width=0.8\columnwidth]{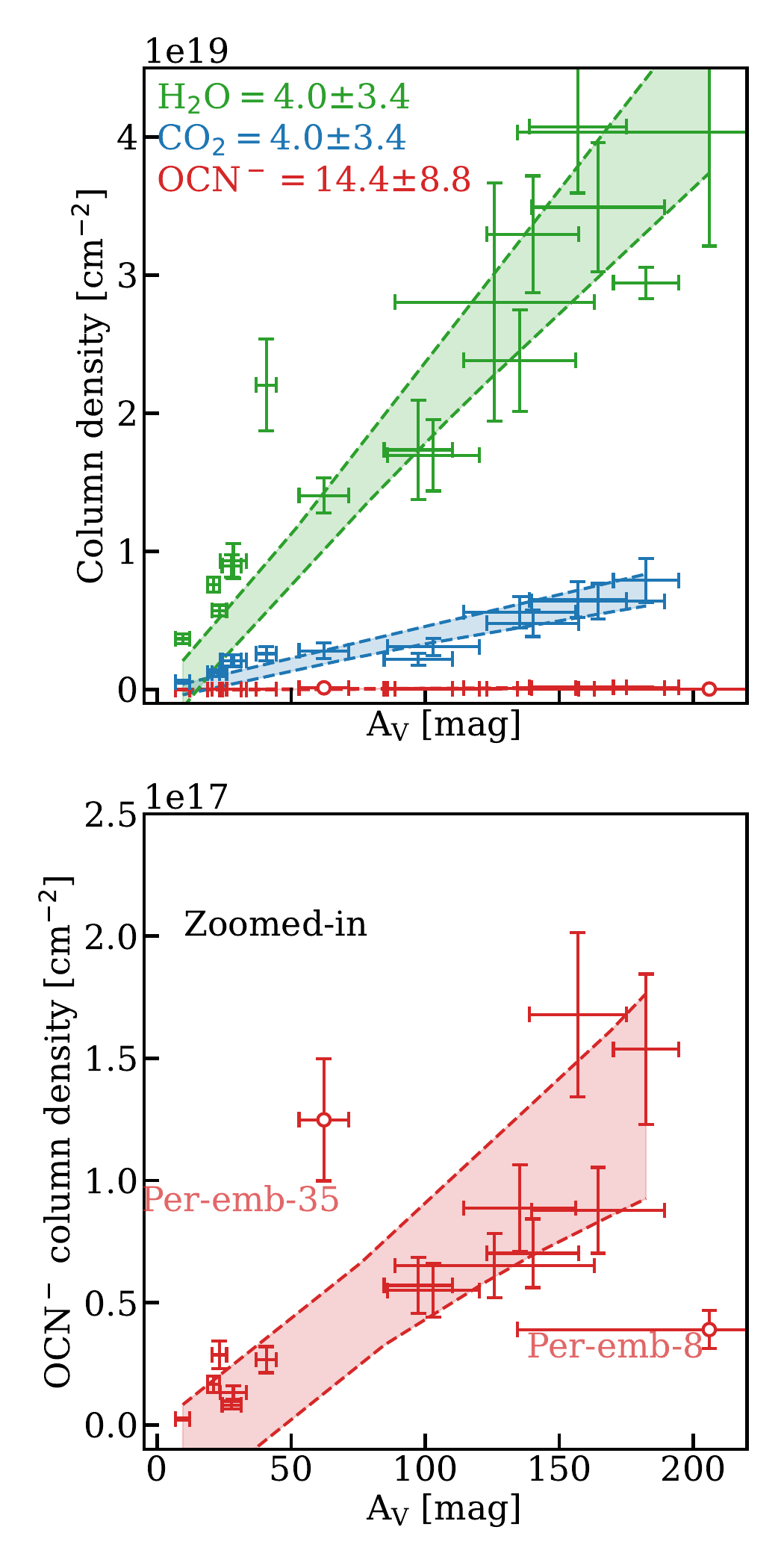}
    \caption{Relation between A$_{\rm V}$ and column densities of H$_2$O (green), CO$_2$ (blue), and OCN$^-$ (red). The lower panel shows a zoomed-in version of the upper panel for OCN$^-$. The two data points of OCN$^-$ shown with a white marker are omitted when fitting the trend. The x-intercept of each molecule is printed in its respective color. The CO$_2$ column densities are taken from \cite{Brunken2024}.}
    \label{fig:Av}
\end{figure}

Figure \ref{fig:Av} presents column densities of H$_2$O, CO$_2$, and OCN$^-$ as a function of A$_{\rm V}$. We find that even up to $A_{\rm V}{\sim}200$ mag, H$_2$O and CO$_2$ show a linear relationship with $A_{\rm V}$ and this relationship still holds despite the warmer environment of the protostellar systems. The OCN$^-$ column densities also show a linear relationship, even though two objects are clear outliers. These two objects are Per-emb-8 and Per-emb-35. The former has an uncertain silicate fitting because the 9.8\,$\mu$m band reaches the noise limit, and therefore the ${\sim}18$\,$\mu$m silicate band was used instead. The latter has particularly a prominent double peak CO$_2$ feature (\citealt{Brunken2025}), and thus its ices are strongly thermally processed (also see Fig. \ref{fig:thermal}). Therefore, we omitted these two sources when fitting the relation for OCN$^-$ as a function of A$_{\rm V}$.

The x-intercept values in Fig. \ref{fig:Av} have large error bars. This is because most of our objects have A$_{\rm V}$ that are larger than ${\sim}40$ and thus extrapolation was needed. As a result, robust conclusions on formation timeline of these molecules are difficult to make with this dataset and clouds with lower A$_{\rm V}$ (e.g. those of \citealt{Smith2025}) are required for such conclusion. Nevertheless, the x-intercept of H$_2$O and CO$_2$ are in agreement within error bars with the ${\sim}2-3$\,mag found toward clouds for these two molecules (\citealt{Boogert2015}). Moreover, the OCN$^-$ x-intercept range goes up to values that are somewhat larger than the range found for the other two molecules, but all three still agree within the uncertainties. To investigate this further, we also considered the relation between CO/H$_2$O and CO$_2$/H$_2$O with OCN$^-$/H$_2$O in Appendix \ref{sec:normalize_H2O}. We did not find any relation between CO$_2$/H$_2$O and OCN$^-$/H$_2$O. This lack of relation together with Fig. \ref{fig:Av} are in agreement with the previous studies which concluded that OCN$^-$ may form later than H$_2$O and CO$_2$ in CO-rich ices.

\section{Conclusions}
\label{sec:conclusion}

In this work we analyzed the absorption bands in the ${\sim}4.4-4.7$\,$\mu$m region for the JOYS+ low-mass protostellar systems. This sample includes Class 0 and Class I objects and extends the pre-JWST results to include objects with lower $T_{\rm bol}$. Our main conclusions are summarized below. 

\begin{itemize}
    \item We detect OCN$^-$ in ices of all our 19 sources. The objects showing a double peak CO$_2$ (15.2\,$\mu$m), a sign of thermal processing, have a factor of ${\sim}2-3$ higher OCN$^-$/CO$_2$ (Fig. \ref{fig:thermal}). Therefore, higher temperatures can enhance OCN$^-$ formation, but only moderately, and detection by itself is not an indicator of thermal processing.
    \item The OCN$^-$, H$_2$O, and CO$_2$ relations with $A_{\rm V}$ (Fig. \ref{fig:Av}) and lack of relation between CO$_2$/H$_2$O and OCN$^-$/H$_2$O tentatively point to formation of OCN$^-$ after H$_2$O and CO$_2$ have already formed in agreement with previous studies.
    \item We find a separation between Class 0 and I objects when considering their OCN$^-$/CO and OCN$^-$/CO$_2$ ratios, with Class 0 systems generally showing larger values. Understanding the origin of this division requires further radiative transfer modeling of Class 0 and Class I objects.   
    \item We also tentatively detect CH$_3$CN, C$_2$H$_5$CN, and N$_2$O in ices of three objects (B1-b, B1-c, and IRAS 2A). The column density ratios of CH$_3$CN, C$_2$H$_5$CN, and N$_2$O with respect to OCN$^-$ are relatively constant (Fig. \ref{fig:ratios_CNs}) pointing to them likely forming in a similar ice environment to OCN$^-$. Out of these three, CH$_3$CN shows the largest ratios with respect to OCN$^-$ ranging between ${\sim}1-10$ and could potentially be an important nitrogen budget in ices.
    \item We find upper limit abundances of ${\sim}2\times 10^{-7}-10^{-6}$ for NO toward five of our objects. These are around one order of magnitude lower than what is expected in ices of the Oph-IRS 48 to explain the gas-phase detection of NO. This likely points to the observed gas-phase NO being a secondary product of another ice species such as N$_2$O. Given similar ice abundances of N$_2$O toward our objects with the NO gas-phase abundance toward Oph-IRS 48, this scenario is likely.
\end{itemize}

In this paper, we show that the JWST sensitivity is key for analysis of ice features toward faint objects with the largest ice column densities. Moreover, the spectral coverage of the entire 4.3-4.8\,$\mu$m region is key for continuum determination and robust column density estimation of OCN$^-$. A larger sample with CO$_2$, OCN$^-$, and H$_2$O measurements are required to further confirm the conclusions made here. This would be particularly useful if measured toward pre-stellar cores. As for the complex cyanides and N$_2$O, future deeper observations (similar to that of B1-c in this work), preferably toward objects with weak gas-phase lines, and laboratory data for ice mixtures of C$_2$H$_5$CN and N$_2$O are needed to make more meaningful deductions.

\begin{acknowledgements}
    We thank K. Pontoppidan and H. Beuther for their insightful comments. This work is based on observations made with the NASA/ESA/CSA James Webb Space Telescope. The data were obtained from the Mikulski Archive for Space Telescopes at the Space Telescope Science Institute, which is operated by the Association of Universities for Research in Astronomy, Inc., under NASA contract NAS 5-03127 for JWST. These observations are associated with programs $\#1186$, $\#1236$, $\#1290$, and $\#1960$. P.N. acknowledges support from the ESO Fellowship and IAU Gruber Foundation Fellowship programs. Astrochemistry at Leiden is supported by funding from the European Research Council (ERC) under the European Union's Horizon 2020 research and innovation program (grant agreement No. 101019751 MOLDISK), the Netherlands Research School for Astron- omy (NOVA), and the Danish National Research Foundation through the Center of Excellence `InterCat' (grant agree- ment No. DNRF150). V.J.M.L.G. acknowledges support by the Spanish program Unidad de Excelencia Mar\'{i}a de Maeztu CEX2020-001058-M, financed by MCIN/AEI/10.13039/501100011033, and by the MaX-CSIC Excellence Award MaX4-SOMMA-ICE. V.J.M.L.G. acknowledges support by the European Research Council (ERC) under the European Union's Horizon 2020 research and innovation program (grant agreement No. 101098309 - PEBBLES). A.C.G. acknowledges support from PRIN-MUR 2022 20228JPA3A ``The path to star and planet formation in the JWST era (PATH)'' funded by NextGeneration EU and by INAF-GoG 2022 ``NIR-dark Accretion Outbursts in Massive Young stellar objects (NAOMY)'' and Large Grant INAF 2022 ``YSOs Outflows, Disks and Accretion: towards a global framework for the evolution of planet forming systems (YODA)''. 
\end{acknowledgements}

%
%

\bibliographystyle{aa}
\bibliography{OCN}

\begin{appendix}

\section{Additional tables}
\label{sec:app_add_tables}

The properties of our objects, the position of spectral extraction, and the rms error on the spectra are given in Table. \ref{tab:obs}. The band strengths for the molecules considered here are presented in Table \ref{tab:band_strengths}. The column densities of different components of CO are given in Table \ref{tab:columns_CO}. It is worth noting that even though, \cite{Boogert2002} multiplied the band strength of the middle CO$_{\rm apolar}$ component by an additional factor of 1.1 to correct for grain shape effects, \cite{Boogert2015}, similar to our work, did not include any additional factors for correction of grain shape effects when calculating the middle CO$_{\rm apolar}$ component column density. In other words, they used the formula $3.5 \tau_{\rm max}/(1.1\times 10^{-17})$, where the denominator is the CO band strength from \cite{Gerakines1995}. In our work, we consistently do not use any correction for the band strength of CO$_{\rm apolar}$ component, both for our measurements and those calculated from the fit parameters given in \cite{Pontoppidan2003} when comparing with the literature values in Fig. \ref{fig:four_panel_text}. The measured $A_{\rm V}$ and H$_2$O column densities are given in Table \ref{tab:columns_water}.

\begin{table}
\centering
\renewcommand{\arraystretch}{1.3}
    \caption{Spectral extraction positions and source properties}
    \label{tab:obs}
    \resizebox{\columnwidth}{!}{\begin{tabular}{@{\extracolsep{1mm}}*{5}{l}}
          \toprule
          \toprule    
        Source & R.A. & Decl. & Class &  Optical depth \\
        &[hh:mm:ss] & [dd:mm:ss] & & rms error\\
        \midrule     

B1-a-N  &  03:33:16.693 &  31:07:55.28 &  I & $1.2 \times 10^{-4}$ \\
B1-a-S  &  03:33:16.677 &  31:07:54.98 &  I & $1.9 \times 10^{-4}$ \\
B1-b  &  03:33:20.334 &  31:07:21.64 &  I & $3.9 \times 10^{-4}$ \\
B1-c  &  03:33:17.893 &  31:09:31.90 &  0 & $3.4 \times 10^{-3}$ \\
EDJ2009183-A  &  03:28:59.304 &  31:15:48.48 &  I & $5.7 \times 10^{-5}$ \\
EDJ2009183-B  &  03:28:59.382 &  31:15:48.48 &  I & $2.6 \times 10^{-4}$ \\
IRAS 2A  &  03:28:55.574 &  31:14:36.90 &  0 & $1.1 \times 10^{-3}$ \\
L1448-mm  &  03:25:38.868 &  30:44:05.66 &  0 & $5.2 \times 10^{-3}$ \\
L1527  &  04:39:53.905 &  26:03:09.69 &  0/I & $5.3 \times 10^{-4}$ \\
Per-emb-8  &  03:44:43.990 &  32:01:35.59 &  0 & $1.2 \times 10^{-3}$ \\
Per-emb-22  &  03:25:22.351 &  30:45:13.18 &  0 & $2.5 \times 10^{-3}$ \\
Per-emb-33  &  03:25:36.316 &  30:45:14.92 &  0 & $1.4 \times 10^{-2}$ \\
Per-emb-35  &  03:28:37.093 &  31:13:30.85 &  0/I & $2.5 \times 10^{-4}$ \\
Per-emb-55-A  &  03:44:43.293 &  32:01:31.33 &  I & $1.9 \times 10^{-4}$ \\
Per-emb-55-B  &  03:44:43.332 &  32:01:31.73 &  I & $1.2 \times 10^{-4}$ \\
S68N  &  18:29:48.140 &  01:16:44.58 &  0 & $2.6 \times 10^{-3}$ \\ 
SMM3  &  18:29:59.308 &  01:14:00.58 &  0 & $3.5 \times 10^{-3}$ \\ 
TMC1-W  &  04:41:12.685 &  25:46:34.42 &  I & $7.7 \times 10^{-5}$ \\
TMC1-E  &  04:41:12.729 &  25:46:34.52 &  I & $7.6 \times 10^{-4}$ \\

\bottomrule
        \end{tabular}}
        \tablefoot{Second and third columns present the right ascension and declination of the aperture center from the NIRSpec cubes used to extract the spectra (i.e., the peak continuum position at 5.251\,$\mu$m).}
\end{table}

\begin{table}
\centering  
\caption{Band strengths}
\label{tab:band_strengths}
\begin{tabular}{lll}
\toprule
\toprule 
Ice species & $A$\,(cm\,molecule$^{-1}$) & Mode\\
\midrule 
CH$_3$CN:CO$_2$ & $1.5 \times 10^{-18}$ & C-N stretching\\
CH$_3$CN:CO & $1.5 \times 10^{-18}$ & C-N stretching\\
CH$_3$CN:H$_2$O:CO$_2$ & $3.3 \times 10^{-18}$ & C-N stretching\\
C$_2$H$_5$CN & $2.9 \times 10^{-18}$ & C-N stretching\\
OCN$^-$ & $1.5 \times 10^{-16}$ & C-N stretching\\
CO  & $1.1 \times 10^{-17}$ & C-O stretching\\
N$_2$O & $5.7 \times 10^{-17}$ & N-N stretching \\
NO & $6.8 \times 10^{-18}$ & N-O stretching \\
H$_2$O (15\,K) & $2.5 \times 10^{-17}$ & libration\\
H$_2$O (150\,K) & $3 \times 10^{-17}$ & libration\\
\bottomrule
\end{tabular}
\tablefoot{Band strengths are taken from \cite{Rachid2022} and \cite{Gerakines2025} for CH$_3$CN mixtures (at 15\,K) and OCN$^-$ (at 10\,K), respectively. The band strength for CO (at 10\,K) is taken from \cite{Gerakines2023}. For C$_2$H$_5$CN we use the value calculated in \cite{Nazari2024_ice}. The band strengths for N$_2$O and NO are taken from \cite{Gerakines2020} and \cite{Fulvio2019}. For the cold and warm water components the band strengths are given in \cite{Mastrapa2009}.}
\end{table}

\begin{table*}
\centering
\renewcommand{\arraystretch}{1.3}
\setlength{\tabcolsep}{12pt} 
\begin{minipage}{0.8\textwidth}
\centering
    \caption{CO ice column densities}
    \label{tab:columns_CO}
    \begin{tabular}{lllll}
          \toprule
          \toprule    
        Source & CO$_{\rm polar}$ & CO$_{\rm apolar}$ & CO$_{\rm blue}$ & Total\\
        & [$\times 10^{17}$ cm$^{-2}$] & [$\times 10^{17}$ cm$^{-2}$]&  [$\times 10^{16}$ cm$^{-2}$] &[$\times 10^{18}$ cm$^{-2}$] \\
        \midrule     

B1-a-N  &  6.4  &  10.0  &  19.0  &  1.8\\ 
B1-a-S  &  5.6  &  8.7  &  17.2  &  1.6\\ 
B1-b  &  36.1  &  33.5  &  26.4  &  7.2\\ 
B1-c  &  70.5  &  22.4  &  60.8  &  9.9\\ 
EDJ2009183-A  &  5.7  &  7.6  &  11.6  &  1.4\\ 
EDJ2009183-B  &  5.1  &  7.3  &  10.9  &  1.4\\ 
IRAS 2A  &  35.4  &  19.3  &  27.6  &  5.7\\ 
L1448-mm  &  25.1  &  7.5  &  21.1  &  3.5\\ 
L1527  &  26.5  &  14.3  &  14.8  &  4.2\\ 
Per-emb-8  &  38.3  &  12.5  &  26.5  &  5.3\\ 
Per-emb-22  &  37.0  &  18.7  &  22.6  &  5.8\\ 
Per-emb-33  &  50.8  &  14.7  &  37.6  &  6.9\\ 
Per-emb-35  &  13.7  &  1.9  &  4.7  &  1.6\\ 
Per-emb-55-A  &  5.1  &  11.3  &  3.4  &  1.7\\ 
Per-emb-55-B  &  5.9  &  10.8  &  2.9  &  1.7\\ 
S68N  &  17.6  &  8.8  &  19.1  &  2.8\\ 
SMM3  &  16.2  &  3.5  &  8.4  &  2.0\\ 
TMC1-W  &  7.1  &  2.8  &  11.5  &  1.1\\ 
TMC1-E  &  8.2  &  3.4  &  14.6  &  1.3\\ 

\bottomrule
        \end{tabular}
        \tablefoot{The uncertainties in the column densities from the fit and the data are estimated as ${\sim}20\%$, while the band strength uncertainties are negligible (${\sim}5\%$, \citealt{Gerakines2023}).}
\end{minipage}
\end{table*}

\begin{table*}
\centering
\renewcommand{\arraystretch}{1.3}
\setlength{\tabcolsep}{12pt} 
\begin{minipage}{0.8\textwidth}
\centering
    \caption{Results of the silicate and water libration mode fitting}
    \label{tab:columns_water}
    \begin{tabular}{llll}
          \toprule
          \toprule    
        Source & $\tau_{\rm peak}$ & A$_{\rm V}$ & H$_2$O\\
        & & [mag] & [$\times 10^{18}$ cm$^{-2}$] \\
        \midrule

B1-a-N  &  $1.5\pm0.3$  &  $28\pm5$ & ${\sim}$9.3\\ 
B1-a-S  &  $1.5\pm0.2$  &  $28\pm4$ & ${\sim}$9.0\\ 
B1-b  &  $2.2\pm0.2$  &  $41\pm4$ & ${\sim}$22.0\\ 
B1-c  &  $6.4\pm0.4$  &  $182\pm12$ & ${\sim}$29.4\\ 
IRAS 2A  &  $5.5\pm0.6$  &  $157\pm18$ & ${\sim}$40.7\\ 
L1448-mm  &  $4.4\pm1.3$  &  $126\pm37$ & ${\sim}$28.0\\ 
L1527  &  $4.9\pm0.6$  &  $140\pm17$ & ${\sim}$33.0\\ 
Per-emb-8  &  $7.2\pm2.5$  &  $206\pm71$ & ${\sim}$40.3\\ 
Per-emb-22  &  $4.7\pm0.7$  &  $135\pm21$ & ${\sim}$23.8\\ 
Per-emb-33  &  $5.8\pm0.9$  &  $164\pm25$ & ${\sim}$34.9\\ 
Per-emb-35  &  $2.2\pm0.3$  &  $62\pm9$ & ${\sim}$14.0\\ 
Per-emb-55-B  &  $0.5\pm0.1$  &  $10\pm3$ & ${\sim}$3.7\\ 
S68N  &  $3.6\pm0.6$  &  $103\pm17$ & ${\sim}$17.0\\ 
SMM3  &  $3.4\pm0.4$  &  $97\pm13$ & ${\sim}$17.4\\ 
TMC1-W  &  $1.1\pm0.1$  &  $21\pm2$ & ${\sim}$7.6\\ 
TMC1-E  &  $1.2\pm0.1$  &  $23\pm3$ & ${\sim}$5.7\\

\bottomrule
        \end{tabular}
        \tablefoot{A$_{\rm V}$ is calculated using the bottom of the silicate feature at ${\sim}9.8$\,$\mu$m ($\tau_{\rm peak}$) and multiplying that by 28.6 and 18.5 for Class 0 and I objects, respectively (see \citealt{Boogert2013} and Appendix G of \citealt{vanDishoeck2025}). The Class 0/I objects are assumed as Class 0 for this conversion.}
\end{minipage}
\end{table*}

\section{Additional plots}
\begin{figure*}
    \centering
    \includegraphics[width=\textwidth]{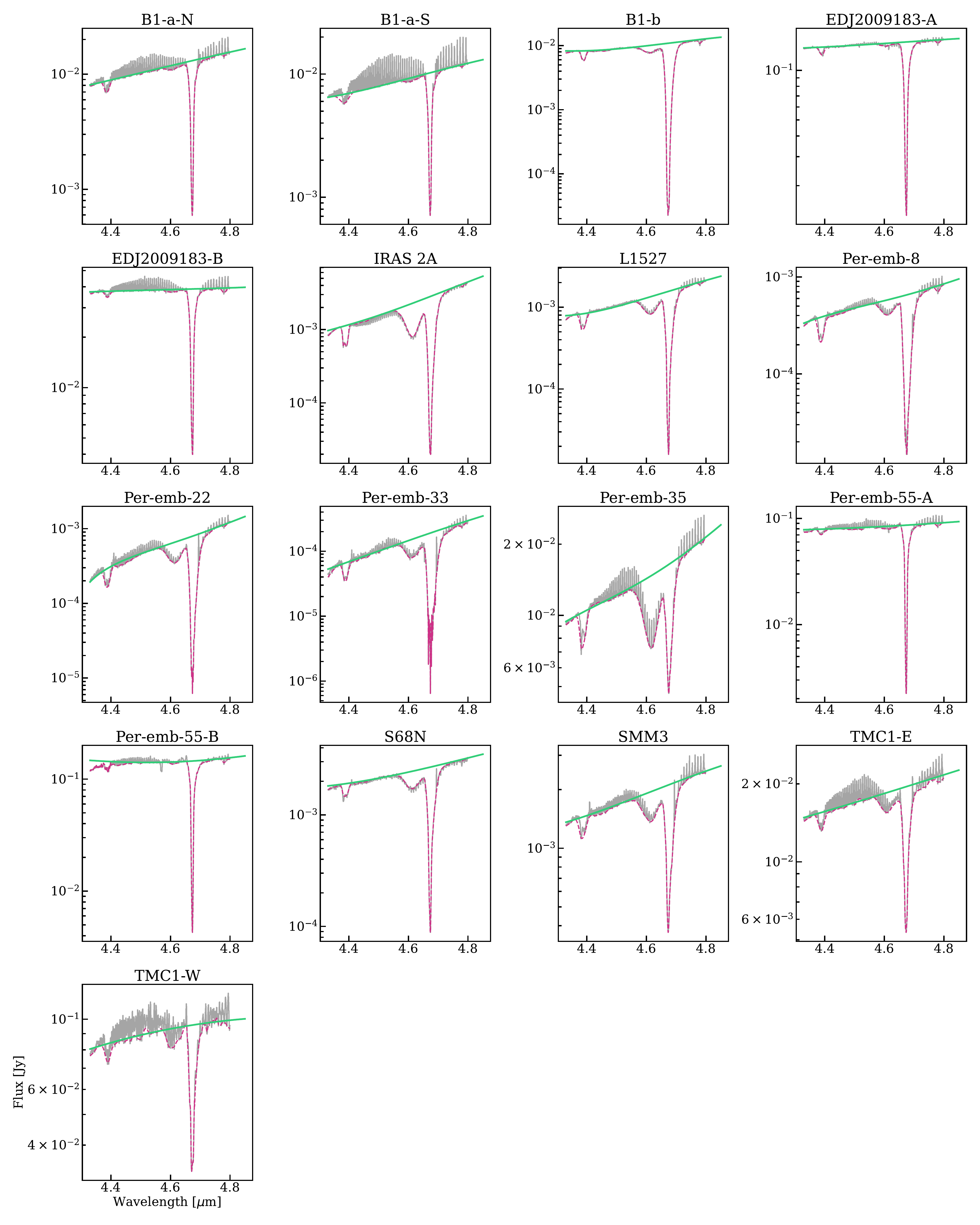}
    \caption{Same as Fig. \ref{fig:spline_text} but for the rest of our sample.}
    \label{fig:splines_app}
\end{figure*}

\begin{figure*}
    \centering
    \includegraphics[width=\textwidth]{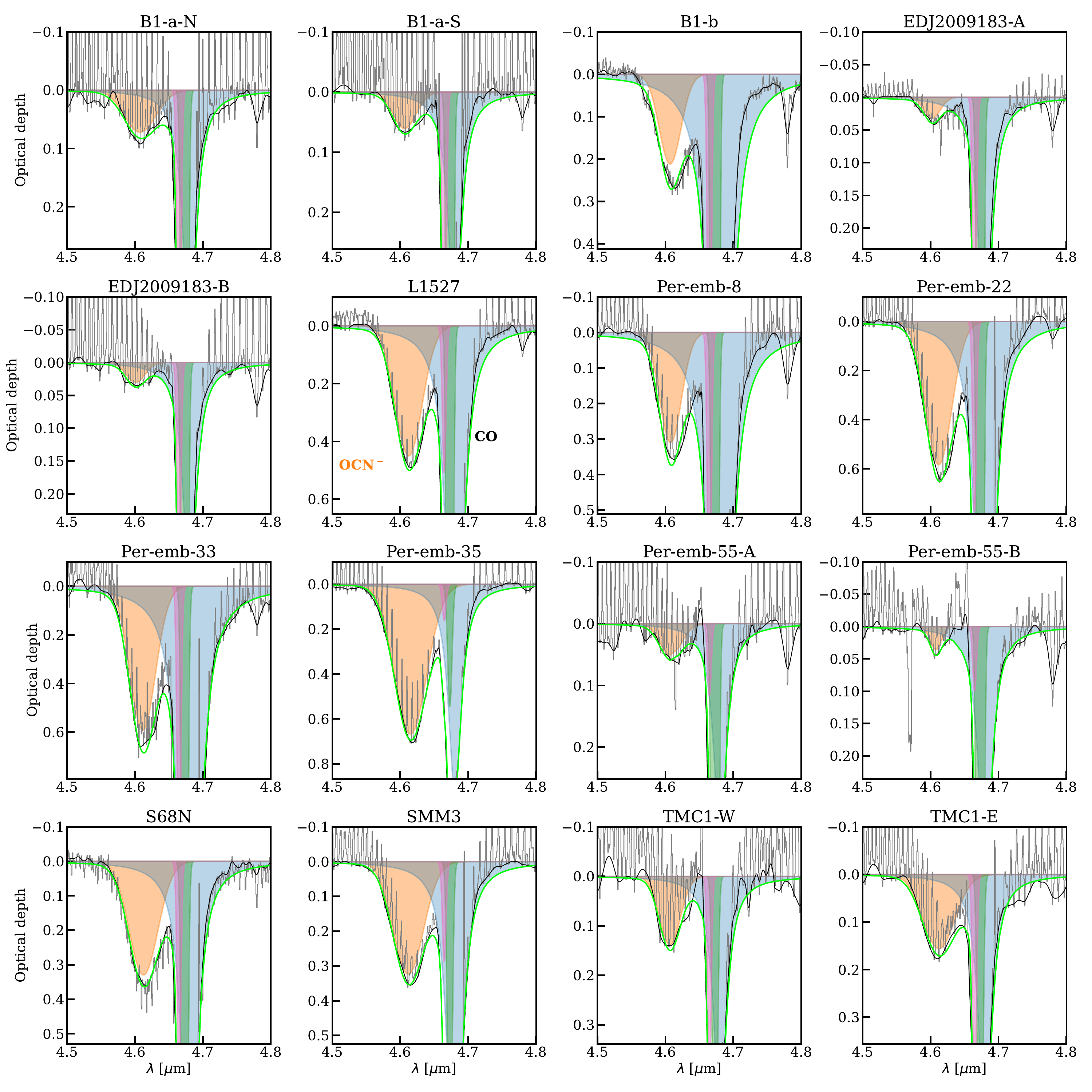}
    \caption{Same as top row of Fig. \ref{fig:OCN_fits} but for the rest of the objects considered.}
    \label{fig:OCN_fits_app}
\end{figure*}

\begin{figure*}
    \centering
    \includegraphics[width=0.9\textwidth]{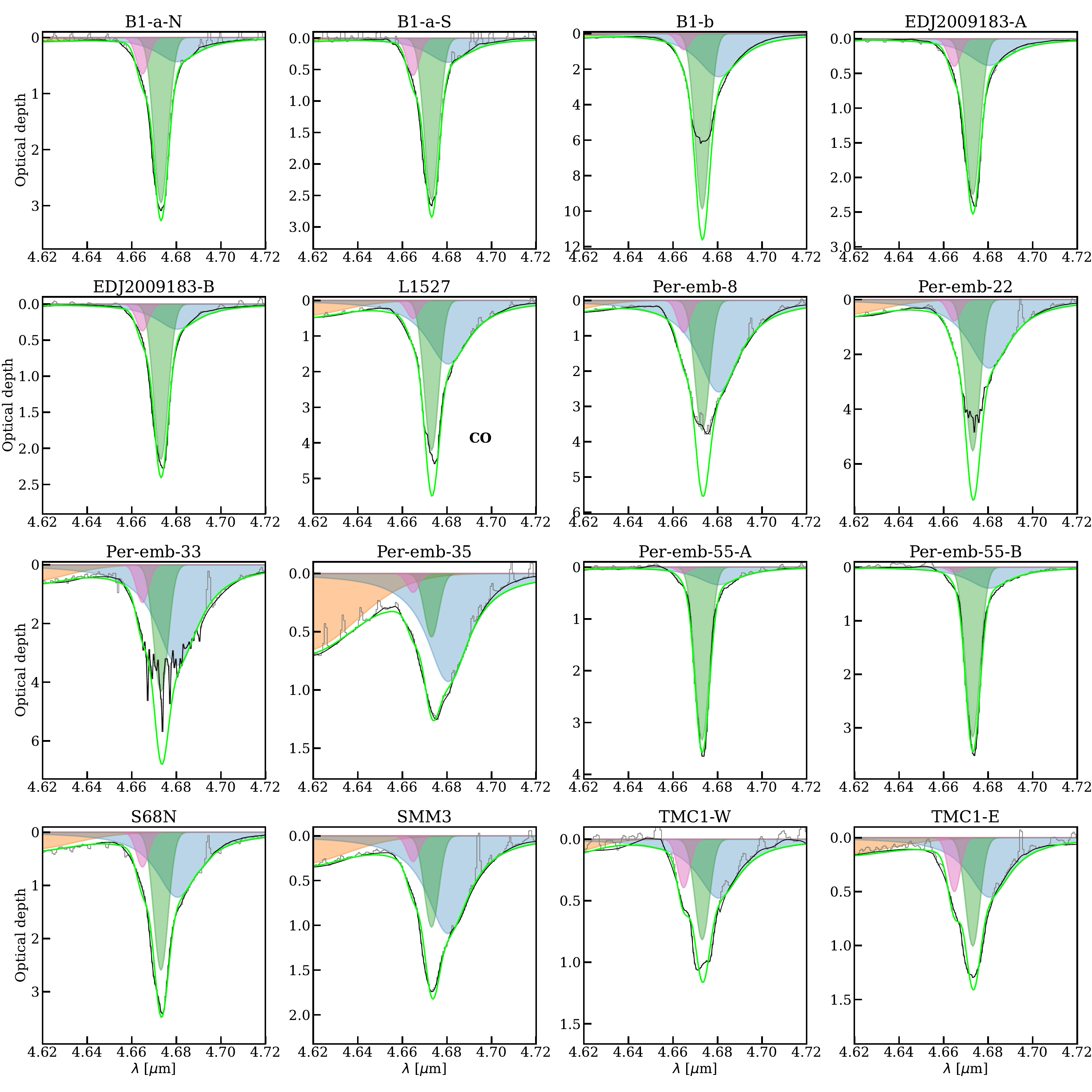}
    \caption{Same as bottom row of Fig. \ref{fig:OCN_fits} but for the rest of the objects considered. The fits that over produce the bottom of the CO band (i.e., B1-b, L1527, Per-emb-8, Per-emb-22, Per-emb-33), are those where only the wings are fitted because the bottom of the CO feature approaches the noise level in flux scale.}
    \label{fig:CO_fits_app}
\end{figure*}

\begin{figure*}
    \centering
    \includegraphics[width=\textwidth]{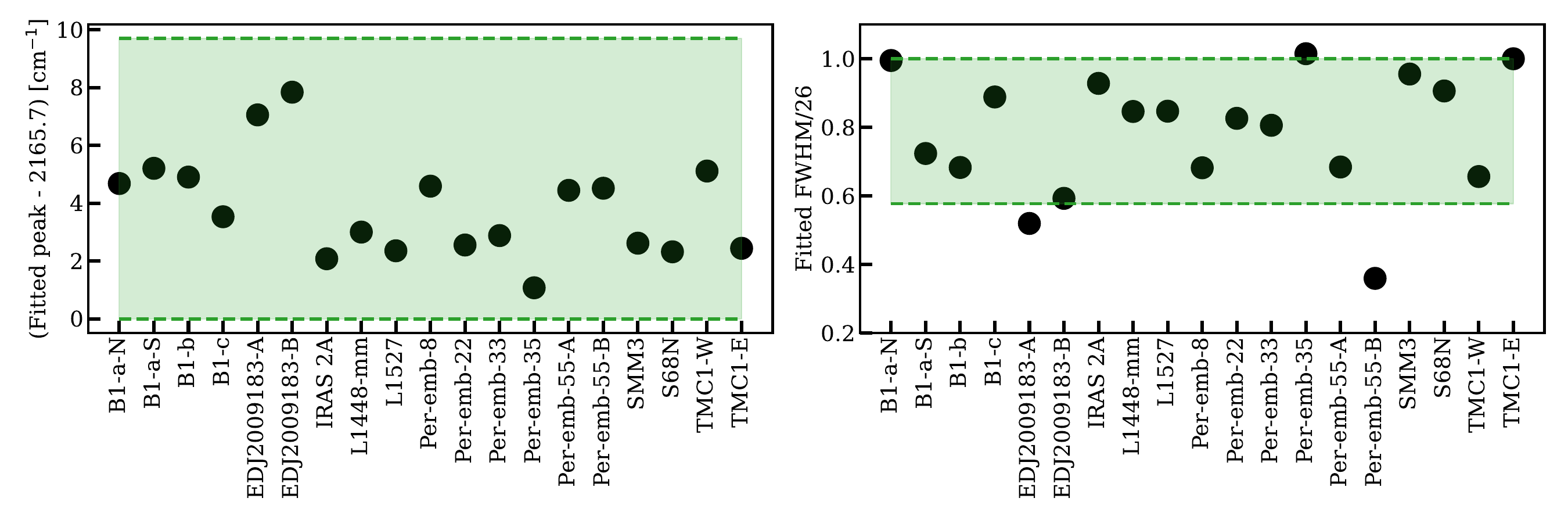}
    \caption{Differences between the fitted peak positions (left) and FWHM (right) for OCN$^-$ in this work and the value suggested by \cite{Broekhuizen2005} for the major component fitted to the 4.6\,$\mu$m band. The shaded green areas show the difference between the major and minor components used in the two-component fit of the 4.6\,$\mu$m band in \cite{Broekhuizen2005}. The fitted values mostly fall within this range.}
    \label{fig:OCN_compare}
\end{figure*}

\begin{figure}
    \centering
    \includegraphics[width=\columnwidth]{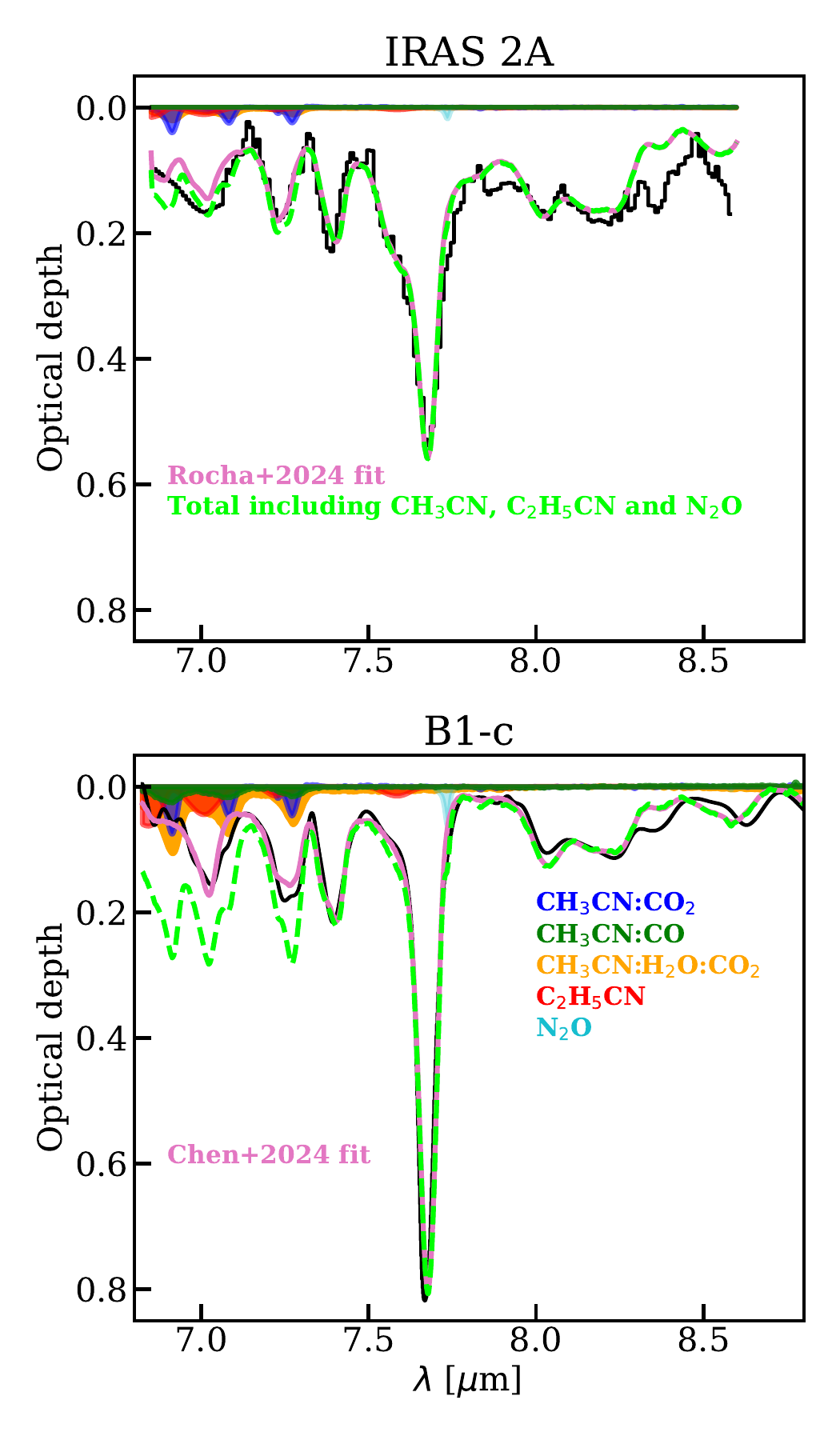}
    \caption{Our CH$_3$CN, C$_2$H$_5$CN, and N$_2$O fits (same colours as Fig. \ref{fig:CH3CN_fits}) on top of the total ice fits from \cite{Rocha2024} and \cite{Chen2024} in pink for IRAS 2A (top) and B1-c (bottom).}
    \label{fig:MIRI_fit}
\end{figure}

\begin{SCfigure*}
    \centering
    \includegraphics[width=0.8\textwidth]{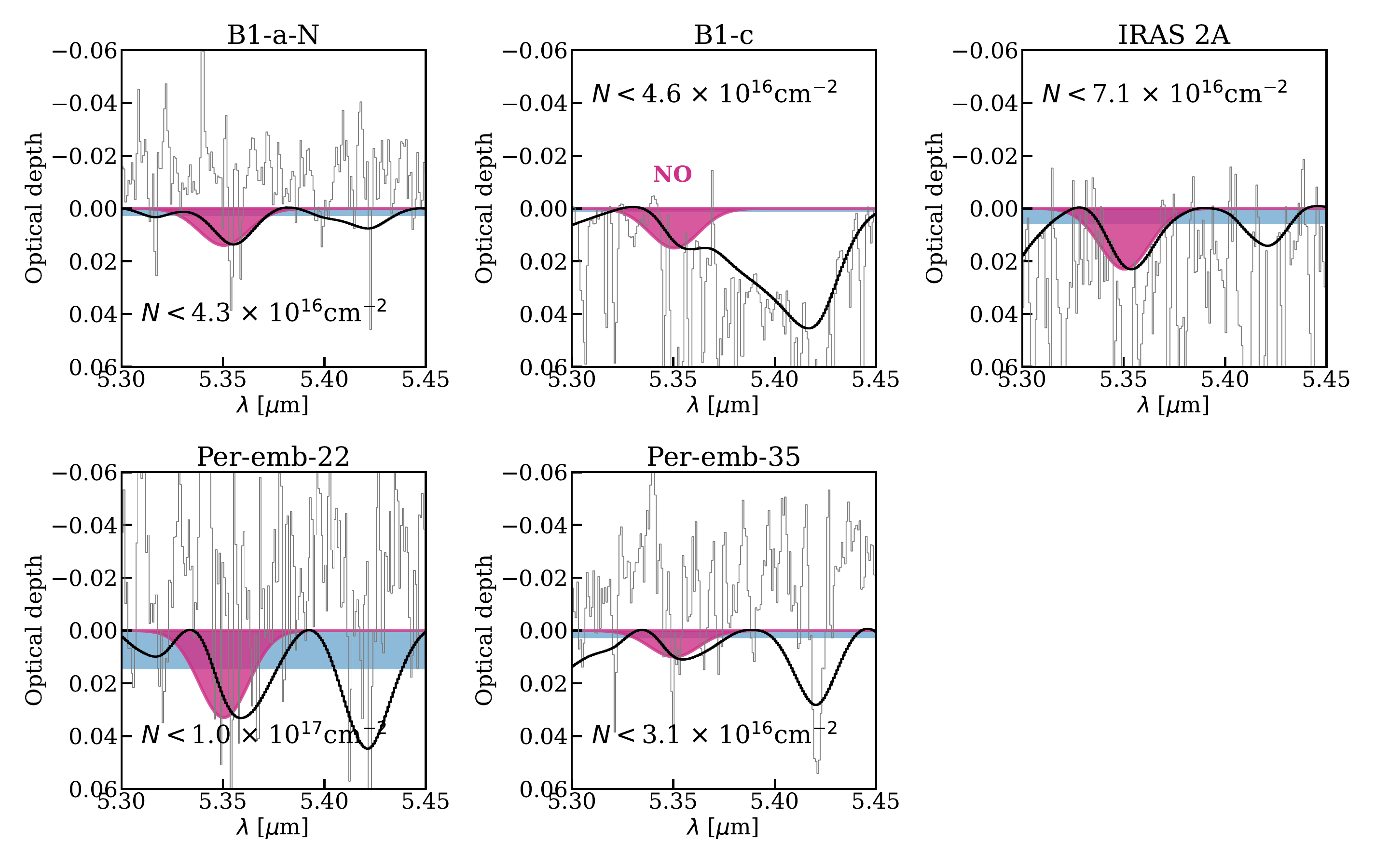}
    \caption{Upper limit fits of NO (shaded pink) toward the objects that showed a tentative absorption feature at 5.35\,$\mu$m. The shaded blue area shows the 3$\sigma$ error on the optical depth in the relevant MIRI wavelength range. Gray shows the data and the black line is the fitted spline.}
    \label{fig:NO}
\end{SCfigure*}

Figure \ref{fig:splines_app} presents the spline and local continuum fitting for the sample. Figure \ref{fig:OCN_fits_app} presents the fits to the OCN$^-$ and CO features combined. This is the continuation of Fig. \ref{fig:OCN_fits} in the main text. Figure \ref{fig:CO_fits_app} presents the same but zooming in on the CO ice band. Figure \ref{fig:OCN_compare} compares the fitted peak positions and FWHMs to the OCN$^-$ ice band with those of suggested by \cite{Broekhuizen2005} with their higher spectral resolution data. Figure \ref{fig:MIRI_fit} presents our fits based on the NIRSpec range on top of the total models found by \cite{Rocha2024} and \cite{Chen2024} for IRAS 2A and B1-c in the MIRI range. Figure \ref{fig:NO} presents the upper limit fits for NO for objects with a tentative absorption feature at ${\sim}5.35\,\mu$m. 

\section{Column densities normalized by H$_2$O}
\label{sec:normalize_H2O}

\begin{figure*}
    \centering
    \includegraphics[width=0.85\textwidth]{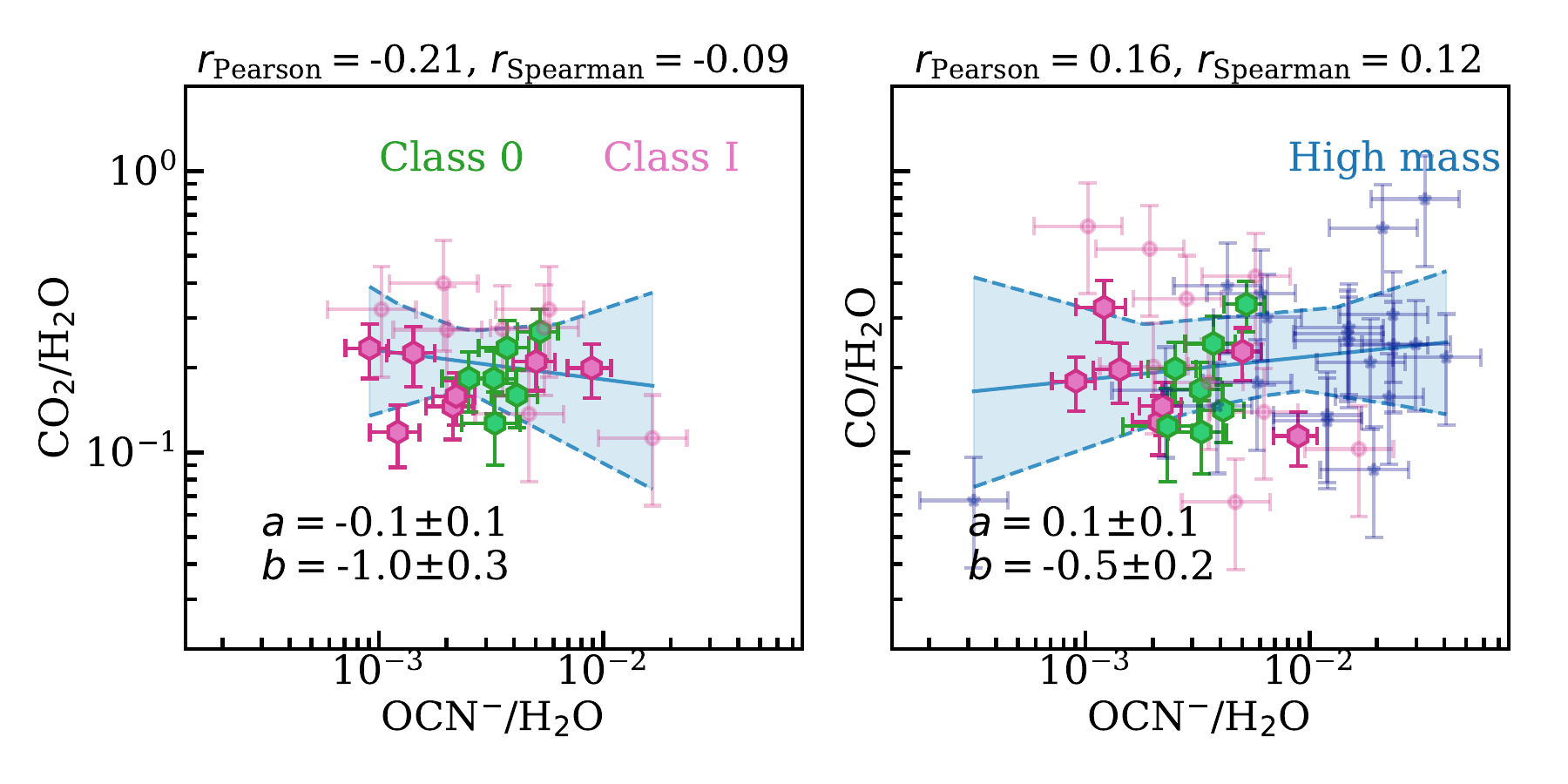}
    \caption{Column density ratios of CO$_2$/H$_2$O and CO/H$_2$O versus OCN$^-$/H$_2$O. The thick hexagons are the JOYS+ JWST results from this work, while the thin circles and stars indicate results from VLT, \textit{Spitzer}, and SpeX for low- and high-mass systems. The OCN$^-$ values are taken from \cite{Broekhuizen2005} and the CO, CO$_2$, and H$_2$O are taken from \cite{Pontoppidan2003}, \cite{Pontoppidan2008}, and \cite{Boogert2008} for low-mass objects. The data for high-mass objects are taken from \cite{Boogert2022}. Green shows Class 0 objects, while pink presents Class Is, and blue presents high-mass. The blue solid line shows a fitted straight line to the data and the blue region the uncertainty in the fit. Pearson's $r$ and Spearman's rank coefficient of $\log_{10}$ of the data are printed on top of each panel. We also fitted a straight line to $\log_{10}$ of the data. The slope ($a$) and intercept ($b$) of each line are printed on the panel.}
    \label{fig:four_panel_text}
\end{figure*}

Another clue on formation of OCN$^-$ may be found by considering the relation between column densities of CO and CO$_2$ as a function of OCN$^-$, all normalized by H$_2$O column densities. Figure \ref{fig:four_panel_text} shows these relations. We also added some of the previous literature data to Fig. \ref{fig:four_panel_text}, where we have considered a $30\%$ uncertainty for non-JWST data points. This is generally larger than what is reported in the original papers and is because those works mainly consider the uncertainty in the fits which are quite small, while the uncertainty in the pre-JWST fluxes are larger which should be reflected in the errorbars. The band strength for all components of CO is assumed as $1.1 \times 10^{-17}$\,cm\,molecule$^{-1}$ for our measurements and is corrected to $1.1 \times 10^{-17}$\,cm\,molecule$^{-1}$ for the literature values to be consistent.

Figure \ref{fig:four_panel_text} shows that there is no relation between OCN$^-$/H$_2$O and CO$_2$/H$_2$O consistent with later formation of OCN$^-$. On the other hand, the relation with CO/H$_2$O seems more complicated. Considering all objects, CO/H$_2$O does not seem to have any correlation with OCN$^-$/H$_2$O. However, excluding the two high-mass objects with the highest CO/H$_2$O points to a negative correlation where objects with more OCN$^-$/H$_2$O are those with less CO/H$_2$O. In Sect. \ref{sec:thermal} we discussed the effect of heating on OCN$^-$ formation. Because high-mass star forming regions likely have higher temperatures the larger OCN$^-$/H$_2$O in Fig. \ref{fig:four_panel_text} for the high-mass objects may be due to enhancement of OCN$^-$ in those warmer regions.

\end{appendix}

\end{document}